\documentclass[pre,aps,10pt,onecolumn,twoside,floats,preprintnumbers,floatfix,superscriptaddress,a4paper,showpacs,showkeys,longbibliography]%
{revtex4-1}

\usepackage{amsmath}
\usepackage{amsfonts}
\usepackage{amssymb}

\usepackage{hyperref}

\usepackage{times}
\usepackage{time}
\usepackage{graphicx}

\usepackage{color}

 \newcommand{\beq}{\begin{equation}}
 \newcommand{\eeq}{\end{equation}}

   \newcommand{\pder}[2]{\frac{\partial {#1}}{\partial {#2}}}

\newcommand{\mum}{\mu \mathrm{m}}

\newcommand{\tbf}[1]{\textbf{#1}}

\def\fig#1{\ref{Fig:#1}}
\def\Fig#1{Fig.~\fig{#1}}

\def\eq#1{(\ref{Eq:#1})}
\def\Eq#1{Eq.~\eq{#1}}
\def\Eqs#1{Eqs.~\eq{#1}}

\begin{document}
\'
\\
\\
\\
\\
\\
\\
\title{Numerical simulation of artificial microswimmers driven by Marangoni flow}
\\
\\
\\
\author{L. Stricker}
 \affiliation{\text{Max Planck Institute for Dynamics and Self-Organization (MPI DS), 37077 G\"ottingen, Germany}}

\begin{abstract}
In the present paper the behavior of a single artificial
microswimmer is addressed, namely an active droplet moving by
Marangoni flow. We
provide a numerical treatment for the main factors playing a role
in real systems, such as advection, diffusion and the presence of
chemical species with different behaviors. The flow field inside
and outside the droplet is modeled to account for the
two-way coupling between the surrounding fluid and the motion of
the swimmer. Mass diffusion is also taken into account. In
particular, we consider two concentration fields: the surfactant
concentration in the bulk, i.e. in the liquid surrounding the
droplet, and the surfactant concentration on the surface. The
latter is related to the local surface tension, through an
equation of state (Langmuir equation). We examine different
interaction mechanisms between the bulk and the surface
concentration fields, namely the case of insoluble surfactants
attached to the surface (no exchange between the bulk and the
surface) and soluble surfactants with adsorbtion/desorption at the
surface. We also consider the case where the bulk concentration
field is in equilibrium with the content of the droplet. The
numerical results are validated through comparison with analytical
calculations. We show that our model can reproduce the typical
pusher/puller behavior presented by squirmers. It is also able to
capture the self-propulsion mechanism of droplets driven by
Belousov-Zhabotinsky (BZ) reactions, as well as a typical
chemotactic behavior.
\end{abstract}

\maketitle

\section{Introduction}
\label{sec:Introduction}
 Unicellular swimmers, e.g., E. coli
bacteria, spermatozoa, or paramecia are typically of a few to
several ten micrometers in size and their swimming velocities are
of the order of one body length per second. Due to their small
size and swimming velocities, the Reynolds number of the flow
involved in their swimming is much smaller than unity. As a
consequence, viscous damping by far dominates over inertia, and the
physics ruling their swimming  is very different from that
applying to swimming in the macro-world \cite{pur77,sta07,gra69}.
While inertia is the dominant propulsion mechanism for swimming in
the macro-world, microorganisms make use of the viscous drag of
the surrounding fluid for their propulsion. Initiated by the
non-trivial nature of viscosity-based swimming and its important
applications in biology, the study of the fundamental physics of
swimming at low Reynolds numbers has become an active field of
research during the past 40 years
\cite{pur77,lig75,lau09,fau06,bra00}.

Several kinds of artificial microswimmers have been produced, based on different mechanisms \cite{dre05,ozi05}, with a
potential for technological applications such as target drug
delivery, removal of pollutants, waste treatment. Examples of
artificial microswimmers consist in active liquid droplets
immersed in a second immiscible liquid and propelled by
Marangoni flow. The underlying idea is to create a non-uniform
surface tension distribution on the droplet surface, through a
non-uniform distribution of surfactants, i.e. molecules of a third
kind dissolved inside the solution and migrating to the surface
\cite{sav73}. To satisfy the local balance of forces at the
interface, a Marangoni stress originates, tangent to the surface,
from the area with a lower surface tension to the one with a
higher surface tension. From this, a convective fluid motion is
set inside and outside the droplet and, therefore the droplet
starts moving.

Such surface tension gradients can rise (a) from chemical
reactions changing either the structure of the surfactant
molecules \cite{han07,thu11,ban12,ban13} or the surfactant
coverage \cite{toy06,toy09} as well as (b) by dissolution or phase
separation phenomena under non-equilibrium conditions
\cite{tha06,mol07}. Intriguing systems have been produced, based
on such principles, such as self-propelled droplets driven by
Belousov-Zhabotinsky (BZ) reactions  \cite{kit11,thu12}, behaving
like non-equilibrium chemical oscillators. Some experiments showed
that nematic liquid crystal droplets, immersed in a solution of
water/ionic surfactant, can develop spontaneous motion
\cite{ped12}. Also in this case, the propulsion mechanism appears
to be a Marangoni flow, originating from a combined effect of the
surfactants and the liquid crystals (i.e. non isotropic molecules)
perpendicularly anchoring to the surface. The details of this
mechanism are still unknown.

Active droplets moving by Marangoni flow can be considered
"squirmers", i.e. swimmers where a tangential velocity of the
surface drives propulsion. Theoretical works on this class of
swimmers have investigated the velocity field generated by
individual squirmers \cite{lig52,bla71}, their hydrodynamic
interactions \cite{ish06,llo10}, and the resulting collective
behavior \cite{ish08}. The majority of this works assumes a
prescribed fluid flow at the surface, instead of deriving it.
However, for self-propelled droplets driven by gradients of mass
concentration or temperature \cite{you59, and89} or by
Belousov-Zhabotinsky (BZ) reactions \cite{kit11}, as well as for emulsion droplets \cite{her14,sch16}
of the kind described in \cite{ped12}, analytical
solutions of the flow profiles are available for simplified cases,
and the stability analysis of their motion has been addressed
\cite{red94,red94b,vel96,yos12,sch13}.

The description of more realistic scenarios requires to account for
a large number of factors, such as the two-way coupling between the flow and the active droplet,
several mutually interacting chemical species, mass exchanges between the droplet's interface
and the surrounding liquid (bulk), chemical reactions. In such cases,
the numerical approach can be a powerful instrument to investigate
the problem at hand.

Several numerical techniques have been developed during the last
decades, to track the motion of interfaces:(i) "interface tracking
methods", where the displacement of the interface is tracked in a
direct-fashion, e.g. through a set of Lagrangian marker points
located on the surface itself (front-tracking \cite{try01}) (ii)
"interface capturing methods", where the motion of the interface
is described in an implicit-fashion, by following the evolution of
an additional function; examples are: immersed boundary
\cite{lai08}, finite elements (FEM) \cite{poz04}, volume of fluid
(VOF) \cite{jam04}, marker-and-cell \cite{tas08}, level set
\cite{xu06,ada03} and hybrid methods such as VOF/level set
\cite{yan07}. Combined approaches between (i) and (ii) have also
been adopted in order to simultaneously address the interface
motion, the flow field around the interface and the surfactant
diffusion on the surface. For instance, Lagrangian points marking
the interface have been used in combination with a finite
difference method for the evolution of the surfactant distribution
on the interface and a boundary integral method for the Stokes
equations in the fluid \cite{egg99}. In the present paper, we use a
 level set approach \cite{osh88,set99_BOOK,set03,osh03_BOOK} with
a continuous surface force formulation \cite{bra92}, in
combination with a flow solver based on a projection method
\cite{cho68} to track the motion of the fluid both inside and
around the droplet. A vast literature is available for level set
methods, including some books \cite{osh03_BOOK,set99_BOOK}. For a
review, we refer the reader to Ref.~\cite{set03}. Level set
methods have proven particularly efficient when singularities and
disconnected bodies are present \cite{set92,gib07,yu07}. They
allow for the treatment of deformable droplets as well as for a
straightforward extension to the case of multiple swimmers.
Several authors have numerically studied biphasic systems with
surfactants using different methods: arbitrary Lagrangian–Eulerian
\cite{yan07}, VOF \cite{jam04}, finite difference/front-tracking
\cite{zah06,mur08,olg13,mur14}, the diffuse-interface or
phase-field method \cite{tei11}. However, to our knowledge, only
few have made use of a level-set formulation, either for the case
of insoluble \cite{tei10,xu03,xu06} or soluble \cite{xu12}
surfactants.

The aim of the present work is to provide a
numerical model that could be adapted to the study of real
artificial microswimmers moving by Marangoni flow, such as
BZ-reaction driven droplets \cite{kit11,thu12,sch13} and
emulsion droplets of the kind described in
\cite{ped12}. The description of these systems requires
to account for the two-way coupling between the droplet's motion and the fluid flow,
as well as multiple mutually interacting concentration fields
(e.g.: empty micelles, filled micelles, liquid crystals, molecular
surfactant \cite{ped12,her14}). We wish to provide here a
numerical treatment for the main phenomena playing a role in such
systems: the flow field, the surfactant advection and diffusion, both in the
liquid bulk and on the surface of the droplet, as well as the
interaction between the bulk and the surface
(adsorption/desorption). The case where a concentration field
dissolved inside the bulk is in equilibrium conditions with a
droplet of the same substance (e.g. liquid crystals in the emulsion droplets of \cite{ped12})
is also addressed.

\begin{figure}
\centering
\includegraphics[width=0.4\textwidth]{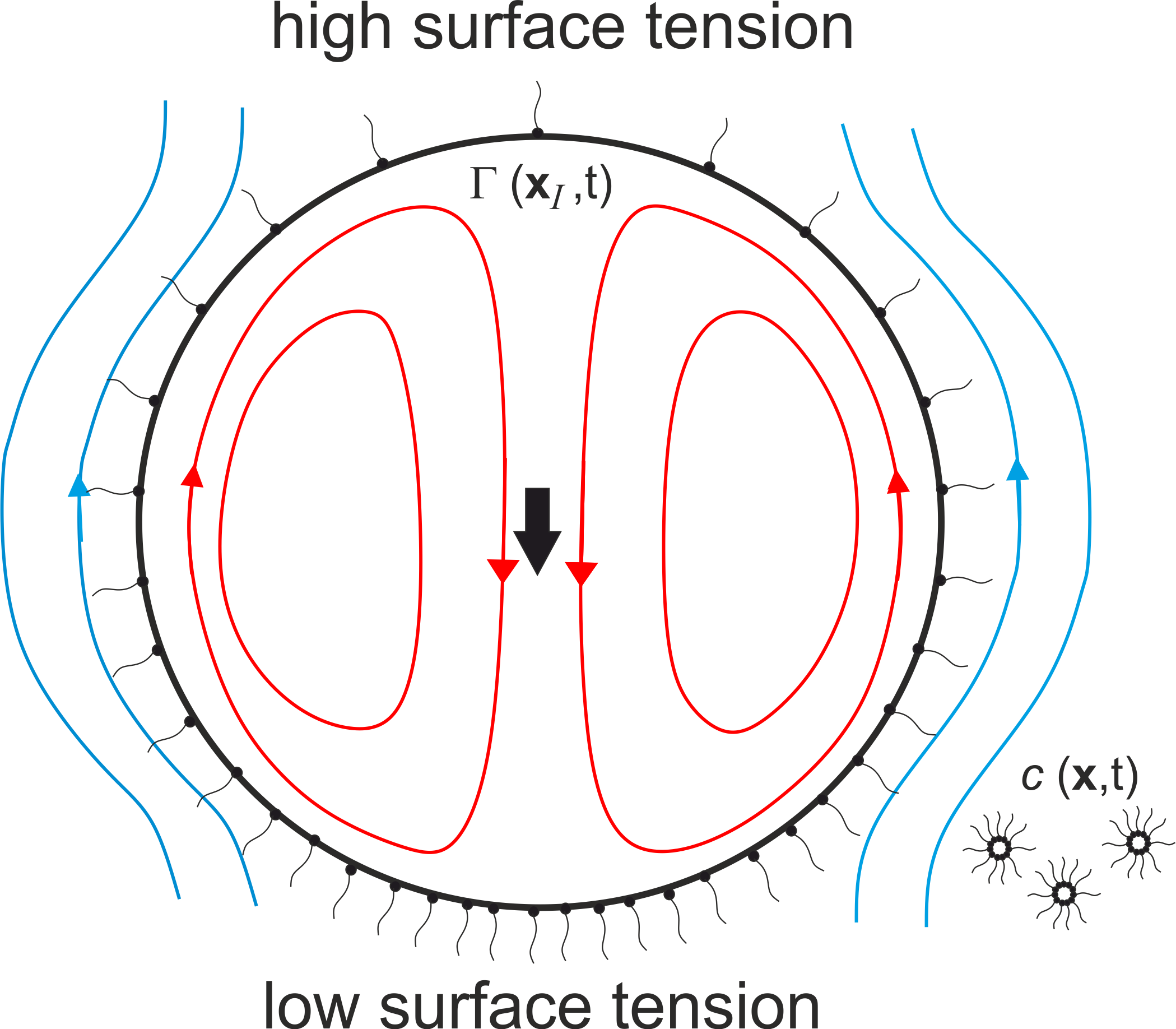}
\caption{Schematic sketch of a droplet moving by Marangoni flow.
A surfactant field is present both on the interface, namely $\Gamma$,
and in the liquid surrounding the droplet (bulk), $c$, in micellar form.
The grey lines (red and light-blue online) represent the flow field inside
(red online) and outside the droplet (light-blue online).
The thick black arrow indicates the direction of the droplet's motion.
}
 \label{Fig:Marangoni sketch}
\end{figure}

In \Fig{Marangoni sketch} we display a simplified sketch of a possible scenario addressed by our code,
where an active droplet moves by Marangoni flow. A surfactant concentration field is present,
both on the surface of the droplet $\Gamma$ and in the bulk $c$. The non-uniform
surface concentration induces a non-uniform surface tension field on the surface.
In particular, a higher concentration of surfactant corresponds to a lower surface
tension, since the surfactants produce a local 'weakness' at the interface.
Through a local force balance, one can see that a tangential flow originates at the
interface - see grey lines (red and light-blue online) - both inside (red online) and
outside the droplet (light blue online). By momentum conservation, the droplet moves
in the direction indicated by the thick black arrow.
The presence of surfactant concentration fields only outside the droplet and on the surface
yields the one-sided nature of the problem. Based on embedded finite
difference schemes, our work presents a novel treatment of such a situation.

\section{Model}
\label{sec:Numerical model}

We consider a 3D geometry, under the assumption of axial symmetry.
The computational domain is fixed and divided in $N_r \times N_z$
cells, of sides $\Delta r = \Delta z = h$, where $r$ is the radial
coordinate and $z$ is the axial coordinate. A level set
description is adopted to track the position of the interface
separating the liquid inside the droplet (oil or liquid crystals)
and the surrounding medium (water). The level set function $\Phi$
is a smooth function defined at the center of each computational
cell. We use a "narrow band" approach \cite{set99_BOOK}, therefore
$\Phi$ is defined only inside a shell $\mathrm{S}_{\Phi}$, with width $2\Delta_{\Phi}$
where $\Delta_{\Phi} = 15h$, located around the interface.
$\Phi$ assumes positive values outside the droplet and negative
values inside. The droplet's interface is
implicitly defined as the set of points where $\Phi=0$. In order
to simplify the numerical treatment, the level set function is
chosen with the particular property of being a signed distance
function. Its absolute value in each point $\textbf{x}$ of the
domain is equal to the shortest distance between $\textbf{x}$ and the
interface $I$ \beq \Phi(\textbf{x}) = S(\Phi(\textbf{x}))
\textmd{min}( | \textbf{x} - \textbf{x}_I |) \, ,\ \ \ \ \forall
\textbf{x}_I \in I \label{eq:signed dist fcn}\, \eeq where
$\textbf{x}_I$ is a point on the interface and $S(\Phi )$ is the
sign of the level set function \beq S(\Phi) =
\begin{cases} 1, & \mbox{if } \textbf{x} \ \textmd{outside \ droplet}  \\ 0, &
\mbox{if } \textbf{x} \ \textmd{on \ surface} \\ -1, & \mbox{if
}\textbf{x} \ \textmd{inside \ droplet}
\end{cases}\, \eeq With this specific choice for the level set
function, the unit normal vector $\textbf{n}$, perpendicular to the
surface at each point, is \beq \textbf{n} = \frac{\nabla
\Phi}{|\nabla \Phi |}\, , \label{eq:normal vector} \eeq and the
local (3D) curvature $\kappa$ can be derived as \beq \kappa = -
\nabla \cdot \textbf{n} \label{eq:local curvature} \eeq

\subsection{Dynamics}
\label{sec:Numerical model, dynamics} We solve the Navier-Stokes
equation inside and outside the droplet, to derive the velocity
field $\textbf{u}(t,\textbf{x})$ and the pressure field
$p(t,\textbf{x})$, as a function of time, $t$ \beq \rho \left[
\pder{\textbf{u}}{t} + \textbf{u} \cdot \nabla \textbf{u} \right]=
- \nabla p + \nabla \cdot [\mu (\nabla \textbf{u} + \nabla
\textbf{u}^T)]+ \delta(\Phi) (-\kappa \sigma \textbf{n} + \nabla_s
\sigma) \, , \label{Eq:momentum eqn}\eeq We close the set of
equations by the incompressibility condition \beq \nabla\cdot
\textbf{u} = 0 \label{Eq:mass conservation}\eeq In the former
equation, $\sigma$ is the surface tension, $\kappa$ is the local
curvature, $\nabla_s = (I - \textbf{n}\otimes \textbf{n}) \nabla$
is the surface gradient and $\delta(\Phi)$ is the surface delta
function. The viscosity $\mu$ and the density $\rho$ of the liquid
are assumed to be constant both inside and outside the droplet. In
the numerical treatment of \Eq{momentum eqn}, we use a smeared
delta function $\delta_{\epsilon}$ instead of $\delta$ \cite{xu12} \beq \delta_{\epsilon}(x) =
\begin{cases}
 \frac{1}{2 \epsilon} \left( 1 + \textmd{cos} \frac{\pi x}{\epsilon} \right),
& \mbox{ if }|x|\leq \epsilon \\ 0, & \mbox{ else }
\end{cases} \label{Eq:delta fn}\eeq In our simulations, we take $\epsilon = 1.5 h$, with $h$ the
length of the side of a computational cell.

The last term in \Eq{momentum eqn} can be seen as a local force
acting only on the interface. It is due to both the surface
tension and its gradient along the interface, the Marangoni
stress. This formulation, the so-called \textit{continuum surface
force formulation} \cite{bra92}, is convenient in order to enforce
the boundary condition for the stresses at the interface, without
explicitly using a jump condition \cite{xu03,tei11}. We note that,
from the physical point of view, the surface tension is defined
only at points $\textbf{x}_I$ belonging to the interface $I$.
However, in order to facilitate the numerical treatment, we adopt
a fictitious surface tension field $\sigma(\textbf{x})$,
defined at the centre of the computational cells, in a
neighborhood of the interface. The value of $\sigma$ at a generic
point $\textbf{x}$ of the computational domain corresponds to the
physical value of the surface tension at the closest point
$\textbf{x}_I$ on the interface and it is derived from an extended
surface concentration field of the surfactants as described below.

The surface tension is related to the local surface concentration
of the surfactants $\Gamma$ through the Langmuir equation of state
\cite{paw96}\beq \sigma(\Gamma) = \sigma_0 + \mathcal{R}T
\Gamma_\infty \log(1-\Gamma / \Gamma_\infty) \label{Eq:Langmuir
eqn}\eeq where $\sigma_0$ is the surface tension for a clean
interface ($\Gamma = 0$), $\Gamma_\infty$ is the saturation
concentration, i.e. the maximum surfactant packing on the surface,
$R$ is the ideal gas constant and $T$ is the temperature,
expressed in Kelvin. \Eq{Langmuir eqn} provides a good description
for low values of the surface concentration ($\Gamma \ll
\Gamma_\infty$) but it has reduced accuracy otherwise. To prevent
unphysical negative surface tension, resulting into an instability
of the interface, we modify \Eq{Langmuir eqn} \cite{mur08} as \beq
\sigma(\Gamma) = \sigma_0 \biggl[\max \biggl(\epsilon_\sigma,\ 1 +
\beta_s \log(1-\Gamma / \Gamma_\infty)\biggr) \biggr]
\label{Eq:Langmuir eqn, cutoff}\eeq where $\epsilon_\sigma = 0.05$
in our simulations and $\beta_s =
\mathcal{R}T\Gamma_\infty/\sigma_0$ is the elasticity number,
quantifying the sensitivity of the surface tension to surface
concentration variations. In our treatment, the fictitious field
$\sigma(\textbf{x})$ is computed cell by cell, from the values of
another fictitious extended field $\Gamma(\textbf{x})$,
representing the local surface concentration of the surfactants at
the closest point of the interface $\textbf{x}_I$. We refer the
reader to Sec.~\ref{sec:Numerical model, surface mass transport}
for the derivation of such a field, from the real surface
concentration at the interface, $\Gamma(\textbf{x}_I)$.

The equations of momentum (\ref{Eq:momentum eqn}) and mass
(\ref{Eq:mass conservation}) conservation are solved with a
projection method \cite{cho68}. We use a standard staggered grid
for the pressure and the velocity fields, with the pressure defined
at the center of the computational cells and the velocity components
on the walls of the computational cells. In particular for each
computational cell, the radial component of the velocity, $u$, is
defined at the center of vertical walls, while the axial component
of the velocity, $v$ is defined at the center of horizontal walls.
At the time instant $n$, the momentum equation (\ref{Eq:momentum eqn}) is numerically
integrated by means of two separate steps, introducing an
intermediate velocity field $\textbf{u}^\star$. The first
intermediate step consists of a \emph{predictor step}, where both
the advective and the diffusive term are treated explicitly.
$\tbf{u}^\star$ is found with a first order Euler method: \beq
\frac{\textbf{u}^\star - \textbf{u}^n} {\Delta t} = - (\textbf{u}
\cdot \nabla \tbf{u})^n + \frac{1}{\rho}(\nabla \cdot \tbf{D} +
\tbf{F})^n \, ,\label{Eq:predictor} \eeq where $\tbf{D} =
\mu6 (\nabla \textbf{u} + \nabla \tbf{u}^T)$ and $ \tbf{F} =
\delta(\Phi) (-\kappa \sigma \tbf{n} + \nabla_s \sigma)$. In order
to prevent instabilities arising from the explicit treatment, the
time step $\Delta t$ is chosen according to the CFL condition, so
that at each time step the displacement of the interface will not
be larger than one grid cell \cite{can12} \beq \Delta t
 = \epsilon_{\tau} \textmd{min}(\tau_{adv},\tau_{visc},\tau_{surf},\tau_{grav})
\label{Eq:CFL condition}\eeq where $\epsilon_{\tau} = 0.25$ and $
\tau_{adv}$, $\tau_{visc}$, $\tau_{surf}$, $\tau_{grav}$ are the
time scales to displace the fluid of one grid cell by advection,
viscosity and capillarity, respectively  \beq \tau_{adv} = \biggl(
\frac{|u_r|_{max}}{\Delta r} + \frac{|u_z|_{max}}{\Delta z}
\biggr)^{-1}
\\, \tau_{visc} = \frac{\rho} {2 \mu}\bigg( \frac{1}{\Delta r^2} +
\frac{1}{\Delta z^2} \bigg)^{-1} \\, \tau_{surf}= \bigg(
\frac{\sigma}{\rho \Delta r^3} \bigg)^{-1/2} \, .\label{Eq:CFL
condition, components}\eeq 
As $\tbf{u}^\star$ is not divergence-free, a \emph{projection
step} is required, to find the real field $\tbf{u}^{n+1}$ at time
$n+1$ \beq \frac{\textbf{u}^{n+1} - \textbf{u}^\star} {\Delta t} =
- \frac{\nabla p^{n+1}} {\rho} \, . \label{Eq:projection} \eeq In
order to calculate the pressure field $p$, we apply the divergence
operator to both sides of \Eq{projection}, imposing the desired
property that the final velocity field is divergence-free, $\nabla
\cdot \tbf{u}^{n+1} = 0 $. Thus we obtain the Poisson equation
\beq \frac{1}{\rho}\nabla ^2 p ^{n+1} = \frac{\nabla \cdot
\tbf{u}^\star}{\Delta t} \, . \label{Eq:Poisson eqn} \eeq This
equation has to be solved subject to the boundary conditions:
$\pder{p}{r}\bigr|_{r=0}=0$ at the z-axis, for symmetry reasons;
at the other limits of the computational domain, we consider a
Dirichlet boundary condition $p=p_\infty$ for open flows and a
Neumann boundary condition $\tbf{n} \cdot \nabla p =0$ for rigid
walls. In the integration of \Eq{predictor}, the spatial
derivatives of $\tbf{u}^n$ in the advective term are calculated by means of a
third-order ENO scheme \cite{osh88}.

\subsection{Kinematics}
\label{sec:Numerical model, kinematics} The value of the level set
function associated to a certain material point must not change
when such a point moves along the trajectory followed by the
interface. Therefore, the level set function is advected by means
of the equation \cite{set03,osh03_BOOK} \beq \pder{\Phi}{t} +
\textbf{u}_{ext}\cdot\nabla \Phi = 0\, , \label{Eq:LS advection
eqn} \eeq in which, $\textbf{u}_{ext}$ is an extended velocity
field, derived from the real fluid velocity field $\textbf{u}$ at
the interface. In the proximity of the interface
$\textbf{u}_{ext}$ corresponds to the real velocity field. In the
far field, a rigid translation of the interface velocity in the
normal direction to the interface itself has been adopted. In
principle, one could use the real velocity field of the fluid,
$\textbf{u}$, for the advection of the level set function
\cite{set03}. However, this method would cause the level set
function to lose its property of being a signed distance function,
which should then be reinforced at every time iteration by mean of
additional numerical manipulations \cite{cho93,sus94,set96,say14},
which are however known to cause mass losses \cite{sus94,sus99}.
To avoid such a problem, we construct an extended velocity field
with the above-mentioned properties and we use it to perform a
rigid-like advection of the level set \cite{ada99}. Several
procedures have been proposed so far to extend a quantity from an
interface, mostly involving the solution of a PDE on the whole
computational domain \cite{ada99,zha96,pen99,cho09}. In the
present paper, we use a different approach. We consider a point
$P$ located in $\textbf{x}_P$, at the center of a computational
cell, where we want to derive the extended velocity field
$\textbf{u}_{ext,P}$. We calculate the location $\textbf{x}_I$ of
the closest point $I$ of the interface, using the signed distance
property of the level set function \beq \textbf{x}_I =
\textbf{x}_P - \Phi_P \textbf{n}_P \, .\label{Eq:closest interface
point}\eeq  We derive the velocity $\textbf{u}_I$ at point $I$ of
the interface with the method described below and we impose
$\textbf{u}_{ext,P} = \textbf{u}_I$. In order to derive
$\textbf{u}_I$, we use a linear interpolation technique based on a
Lagrangian multiplier formulation \cite{can12}. In particular, we
assume that the approximated velocity $\tilde{\textbf{u}}$ at a
generic location $\textbf{x}$ in the neighborhood of point $I$
could be written as \beq \tilde{\textbf{u}}(\textbf{x}) \simeq
\textbf{u}_I + \nabla {\textbf{u}_I} \cdot (\textbf{x} -
\textbf{x}_I) \\, \label{Eq:linear interpolant} \eeq where
$\textbf{u}_I$ is the velocity at point $I$, still unknown. We
consider a portion of the computational domain, namely a square of
$N_s$ x $N_s$ cells around $\textbf{x}_I$ (see the area delimited by the
grey dashed line in \Fig{Extrap velocity, sketch}). We then write the
Lagrangian function $\mathcal{L}$ as \beq \mathcal{L} =
\mathcal{L}_0 + \lambda \mathcal{Q} \, ,\label{Eq:Lagrangian
function} \eeq where $\lambda$ is another unknown parameter,
$\mathcal{Q} = \bigtriangledown \cdot \textbf{u}_I$ represents the
constraint of the velocity field being divergence-free at the
interface and \beq \mathcal{L}_0 = \sum_{i}^{N_s \times N_s}
(\tilde{\textbf{u}}_i - \textbf{u}_i)^2 , \label{Eq:Lagrangian
function, sum square errors} \eeq with $i$ denoting the different
points and $\tilde{\textbf{u}}_i$ expressed by substituting the
values $\textbf{x}_i$ inside \Eq{linear interpolant}. We minimize
the Lagrangian function $\mathcal{L}$ by solving the system \beq
\nabla\mathcal{L} = 0 , \label{Eq:grad(L) = 0} \eeq where the
unknowns are $\textbf{u}_I$,
$\left(\pder{\textbf{u}}{r}\right)_I$,
$\left(\pder{\textbf{u}}{z}\right)_I$ and $\lambda$. Hence we find
the desired interpolated value of the velocity at the interface
$\textbf{u}_I$ and we rigidly transport it in $\textbf{x}_P$,
finding $\textbf{u}_{ext,P}$ (see \Fig{Extrap velocity, sketch}).
The present method is more precise
respect to a standard bilinear interpolation, as it allows for the
possibility to add the divergence-free
constraint of the values of the interpolated velocity at the interface. \\

\begin{figure}
\centering
\includegraphics[width=0.4\textwidth]{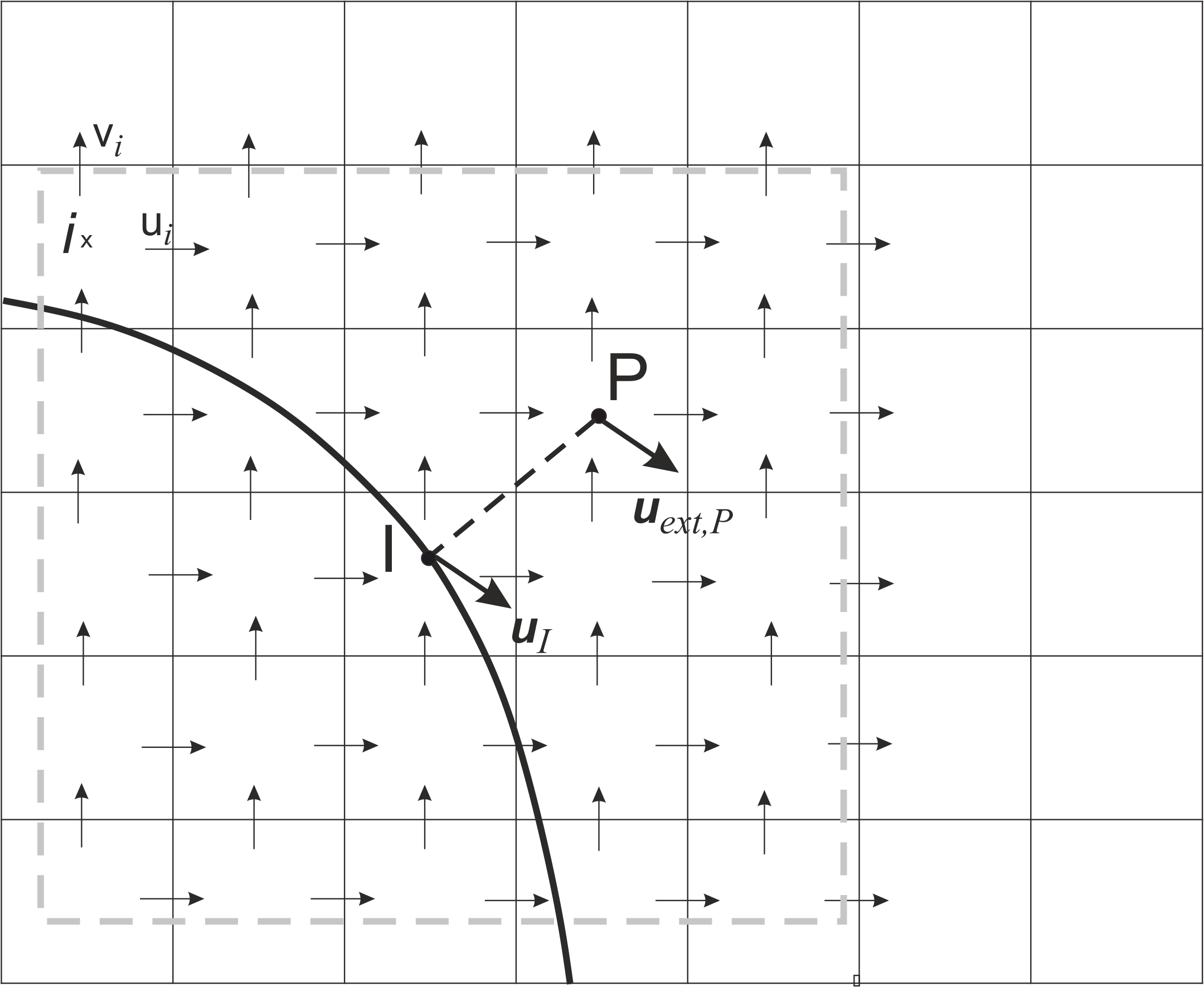}
\caption{Sketch of the points used to derive the extrapolated velocity at
point $P$, $\textbf{u}_{ext,P} = \textbf{u}_I$, to be used for the advection of the
level set function. The interface velocity $\textbf{u}_I$ is computed
based on the calculated velocities in the cells enclosed by the dashed grey line,
whose radial and axial components, $u$ and $v$, are depicted by the horizontal and vertical arrows
respectively. } \label{Fig:Extrap velocity, sketch}
\end{figure}

Once we have derived the extended velocity field, the advection of
the level set function is performed by integrating in time \Eq{LS
advection eqn} with a second-order Adam-Bashforth scheme \beq
\frac{\Phi^{n+1}-\Phi^n}{\Delta t}=-\frac{3}{2}(\textbf{u} \cdot
\nabla \Phi)^n + \frac{1}{2}(\textbf{u}\cdot \nabla \Phi)^{n-1} \,
, \label{Eq:LS advection eqn, Adam Bashforth} \eeq in which the
superscript indicates the time instant. The spatial derivatives of
$ \Phi$ are computed with a fifth-order WENO scheme \cite{jia00}.

After advecting the level set, a correction of $\Phi$ may still become
necessary in order to enforce the signed-distance property. This
manipulation, called \emph{reinitialization}, must not displace
the interface and it can be performed in different ways
\cite{cho93,sus94,set96,say14}. However, due to undesired
interface displacements, the reinitialization step is known to
cause mass loss. Several solutions have been proposed for this
problem \cite{sus94,sus99,rus00,enr02,cho01,che07}. Here, the use
of an extended velocity field of the kind described above for the
advection of the level set, allows us to consistently reduce the
number of required reinitialization steps
\cite{cho93,ada95,cho09,say14}. Since $|\nabla \Phi| = 1$ for a
signed distance function, we perform a reinitialization when
$|\nabla \Phi|$ differs from 1 more than $\epsilon = 10^{-4}$. As
a direct calculation of the zero level set \cite{cho93} is
computationally expensive, in the present work we adopt a modified
version of the method of Ref. \cite{sus94}, solving the equation
\beq \begin{split} \pder{\Phi}{\tau} + S(\Phi ^0)(1-|\nabla \Phi |) &= 0\\
\Phi(\textbf{x},0) &= \Phi^0(\textbf{x},0) \, ,
 \end{split} \label{Eq:LS reinit eqn} \eeq until steady state is reached.
Here $\tau$ is an artificial time and $\Phi^0$ is the level set
function calculated with \Eq{LS advection eqn}.

\Eq{LS reinit eqn} is integrated in the pseudo-time $\tau$ with
the third-order Runge-Kutta scheme described in \cite{shu88},
consisting of a linear combination of Euler integration steps:
\beq \begin{split} \Phi^{(1)} &= \Phi ^t - \Delta \tau L\Phi^t \, ,\\
\Phi^{(2)} &= \frac{3}{4} \Phi^t +
\frac{1}{4} \Phi^{(1)} - \frac{1}{4} \Delta \tau L\Phi^{(1)} \, ,\\
\Phi^{t+1} &= \frac{1}{3} \Phi^t + \frac{2}{3} \Phi^{(2)} -
\frac{2}{3} \Delta \tau L\Phi^{(2)} \, ,
\end{split} \label{Eq:LS reinit eqn, Adams-Bashforth} \eeq
with the superscript $t$ denoting the pseudo-time step,
$L\Phi^{(k)} = S(\Phi ^0) \bigl(|\nabla \Phi^{(k)}| - 1 \bigr)$.
In order to further reduce the mass loss associated to the
reinitialization, particular care has been taken in the
calculation of $\nabla \Phi$ at points adjacent to the interface,
when solving \Eq{LS reinit eqn}. Here the spatial derivatives of
$\Phi$ are computed using the "subcell fix" method described in
Refs. \cite{rus00,can12}. With this method, the spatial
derivatives of $\Phi$ at point \textbf{x} are calculated with a
fifth-order WENO scheme, when \textbf{x} is far from the
interface, and with a third-order modified ENO scheme when
\textbf{x} is close to the interface. The modified ENO schemes
uses an asymmetric stencil, which includes the closest point of
the interface along the direction of the partial derivative that
we are calculating. Such an interface point corresponds to a level
set value known and null; its position is retrieved by cubic
interpolation along the considered direction. This method is
conceptually equivalent to solving \Eq{LS reinit eqn} on two
separate domains, inside and outside the droplet, with Dirichlet
boundary condition $\Phi = 0$ at the interface. The
above-mentioned procedure is essential to guarantee the upwind
nature of the scheme, i.e. to avoid transfer of information across
the interface. It makes use of the fact that the reinitialization
equation is a hyperbolic equation, such that its characteristics
always propagate from the interface to the liquid \cite{can12}.

\subsection{Mass transport processes}
\label{sec:Numerical model, mass transport}
\subsubsection{Surface concentration field $\Gamma$} \label{sec:Numerical model, surface mass
transport}The evolution of the surfactant concentration on the
surface is described by an advection-diffusion equation. The
space-wise implicit nature of the level set formulation, does not
allow for an explicit treatment of a surface concentration field
located at the interface. Therefore, in line with the spirit of
the level-set method, we consider a fictitious field $\Gamma$,
defined at the center of the cells of the computational domain in
a neighborhood of the interface (within a shell
$\mathrm{S}_{\Gamma} = 2 \Delta$ with $\Delta = 10 h$). This
extended field is built in such a way that its interpolated values
correspond to the surface concentration at the interface. In
particular, at each point $P$ of the computational domain, the
value of $\Gamma$ represents the value of the surface
concentration at the closest point of the interface, $I$.
Following a derivation outlined in \cite{sto90,xu03}, we rewrite
the local mass conservation equation as \beq \pder{\Gamma}{t} +
\textbf{u}_{ext} \cdot \nabla \Gamma + \Gamma \nabla \cdot
\textbf{u}_{ext} - \textbf{n} \cdot (\nabla \textbf{u}_{ext}
\textbf{n}) \Gamma = D_s \nabla^2_s \Gamma + j \label{Eq:surface
conc adv-diff eqn}\eeq where $D_s$ is the surface diffusivity
 and $\nabla^2_s \Gamma= \nabla^2 \Gamma - \textbf{n}
\cdot (\nabla \nabla \Gamma)\textbf{n} - \kappa \textbf{n} \cdot
\nabla\Gamma$ is the surface Laplacian. The term $- \textbf{n}
\cdot (\nabla \textbf{u} \textbf{n}) \Gamma$ accounts for the
change in the surface concentration due to the change in the local
curvature. The advection velocity of the level-set function
$\textbf{u}_{ext}$ (see \Eq{LS advection eqn}) describing the
velocity at the interface, is used for the advection of the
surface concentration field; the term $j$ describes the net
exchange of surfactant with the bulk field \cite{tei11} \beq j =
r_a (\Gamma_\infty - \Gamma) c_I - r_d \Gamma \label{Eq:coupling
term surface/bulk}\eeq Here $r_a$ and $r_d$ are the adsorption and
desorption rate, respectively, $\Gamma_\infty$ is the surface
saturation concentration and $c_I$ is the bulk concentration at
the interface. The concentration field in the bulk $c$ is defined
at the center of the computational cells in the liquid outside the
droplet. Unlike $\Gamma$, $c$ is not a fictitious field and its
values indicates a local concentration at the point where they are
defined. Hence, for the calculation of $j$ by \Eq{coupling term
surface/bulk} at a generic point $P \in \mathrm{S}_{\Gamma}$ of
the domain, $\Gamma(\textbf{x}_{P})$ gives the desired value of
the surface concentration at the closest point of the interface
$I$ by construction, but the corresponding bulk concentration
$c_I$ still needs to be calculated. To this aim, we adopt a
procedure based on a Lagrangian multiplier method, similar to the
one we used to derive the extended velocity field
$\textbf{u}_{ext}$ (see \Eqs{LS advection eqn}-\ref{Eq:grad(L) =
0}). We first derive the position of the closest interface point
I, by \Eq{closest interface point}. We assume that the
approximated bulk concentration $\tilde{\tilde{c}}$ at a generic
location $\textbf{x}$ in the neighborhood of point $I$ could be
written as \beq \tilde{\tilde{c}}(\textbf{x}) \simeq c_I + \nabla
c_I \cdot (\textbf{x} - \textbf{x}_I) + (\textbf{x} -
\textbf{x}_I)^T \cdot \mathbf{D}_2(c) \cdot (\textbf{x} -
\textbf{x}_I) \\, \label{Eq:cubic interpolant} \eeq where
$\mathbf{D}_2$ denotes the Hessian matrix. The Lagrangian function
is defined as \beq \mathcal{L} = \mathcal{L}_0 = \sum_{i}^{N_p}
(\tilde{\tilde{c}}_i - c_i)^2 ,\label{Eq:Lagrangian function, bulk
conc} \eeq with $N_p$ the number of valid points, i.e. points
where $\Phi > 0$ and the field $c$ has a physical meaning. The
$N_p$ points are chosen based on a minimum distance criterium from
the point $I$ itself, i.e. we take the $N_p$ closest points to $I$
(see \Fig{Extrap bulk conc, sketch}).
The fitting function $\tilde{\tilde{c}}$ is
found again by solving the system $\nabla\mathcal{L} = 0$. Hence
we know $c_I$. We remark that this procedure was adopted in order
deal with both the one-sided nature of the problem and the stiff gradients
of the bulk concentration field, that may originate close to the interface.
We verified that, in a case like this, where one
wishes to find a one-sided interpolation respect to an interface,
this method gives better results respect to a standard bicubic
interpolation based on a 16 points stencil as it requires a lower
number of points $N_p$. In our simulations we took $N_p = 6$ and
we checked that this was enough to provide good results.

\begin{figure}
\centering
\includegraphics[width=0.3\textwidth]{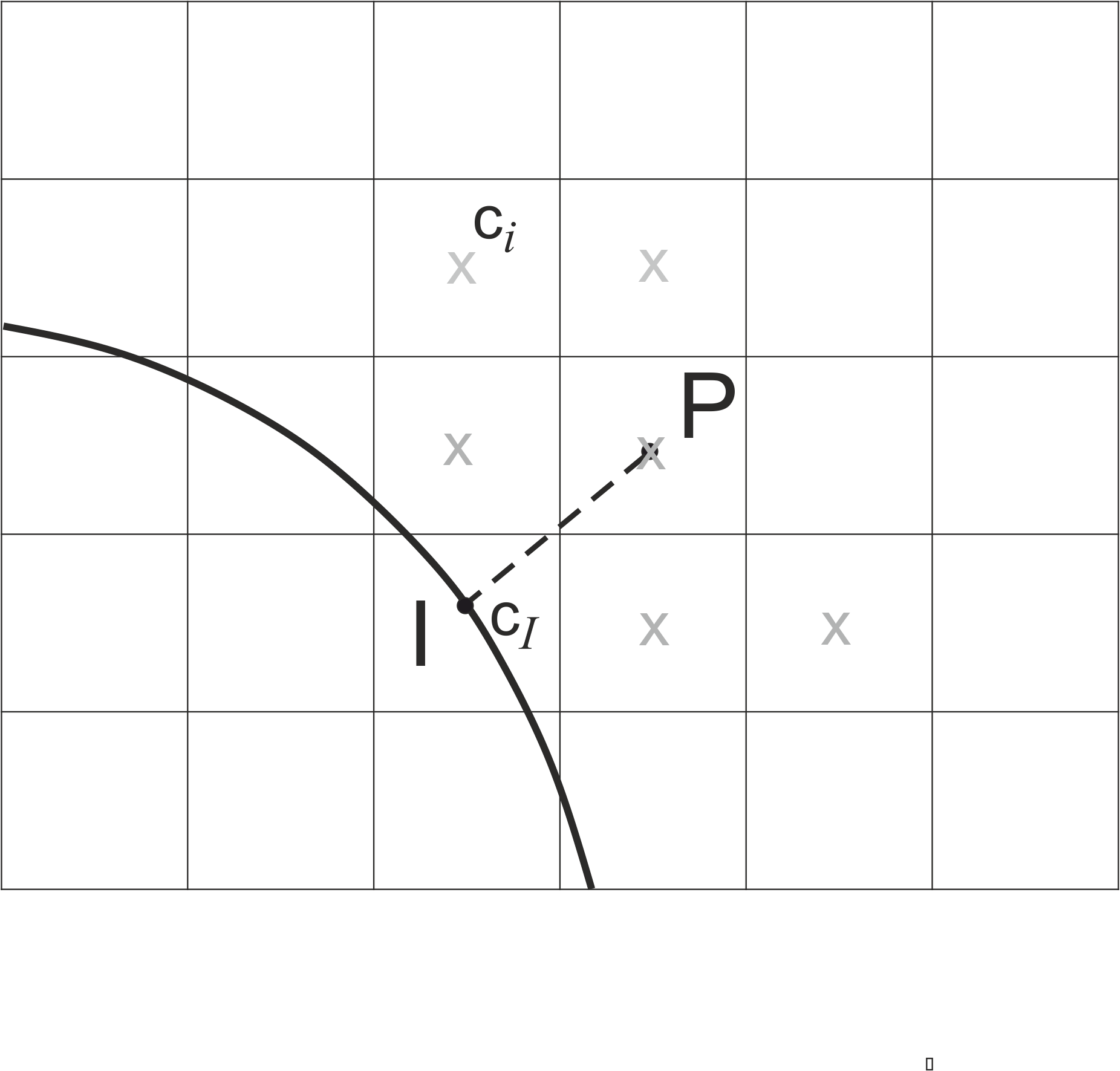}
\caption{Sketch of the points used to infer the interpolated bulk concentration at
point $I$, $c_I$, required for the calculation of the source term $j$ at point $P$ (see \Eq{coupling term
surface/bulk}). The grey crosses represent the values of the bulk concentrations used to derive $c_I$. }
 \label{Fig:Extrap bulk conc, sketch}
\end{figure}

\Eq{surface conc adv-diff eqn} is integrated in a semi-implicit
fashion, by solving the linear system \beq \frac{\Gamma^{n+1} -
\Gamma^n}{\Delta t} = \frac{3}{2}R^n - \frac{1}{2}R^{n-1} + D_s
\bigg(\frac{1}{2} \nabla^2_s \Gamma^{n+1} + \frac{1}{2} \nabla^2_s
\Gamma^n \bigg) \label{Eq:surf conc adv-diff eqn, Adam Bashforth}
\eeq where $R^n = [-\textbf{u} \cdot \nabla \Gamma + \textbf{n}
\cdot (\nabla \textbf{u} \textbf{n}) \Gamma + j]^n$. The
superscripts $n$ and $n-1$ indicate the present and the previous
time instants, respectively, where all the quantities are known,
while $n+1$ is referred to the next time step, therefore to
unknown quantities. The spatial derivatives of $\Gamma$ in the
explicit advective term are computed with a fifth-order WENO
scheme \cite{jia00}. For the derivatives in the diffusive implicit
term, we adopt a standard second order centered difference
formulation. At the borders of the area $\mathrm{S}_{\Gamma}$ we
impose a Neuman boundary condition $\textbf{n} \cdot \nabla\Gamma
= 0$. The linear system deriving from the discretization of
\Eq{surf conc adv-diff eqn, Adam Bashforth} is solved by means of
a conjugate gradient method. We chose to use a semi-implicit
solver in time, because it does not give rise to additional
constraints on $\Delta t$ respect to those imposed by the CFL
condition (see \Eqs{CFL condition}-\ref{Eq:CFL condition,
components}). Thus it allows for larger time steps respect to an
explicit method.

Similar to the advection of the level set function, there are
discretization errors in the advection of the extended surface
tension field. We hence need to correct the advected extended
field $\Gamma$ to ensure that the concentration values at the
interface are properly transported along the characteristics. To
this aim we adopt a scheme in the same spirt of the one described
in \Eqs{LS reinit eqn}-\ref{Eq:LS reinit eqn, Adams-Bashforth} for
the reinitialization of the level set function, i.e. we solve the
following equation until steady state is reached \cite{asl04}
\beq \begin{split} \pder{\Gamma}{\tau'} + S(\Phi ^0)(\textbf{n} \cdot \nabla \Gamma) &= 0\\
\Gamma(\textbf{x},0) &= \Gamma^0(\textbf{x}) \, ,
 \end{split} \label{Eq:extend
surf conc}\eeq where $\tau'$ is an artificial time and $\Gamma^0$
is the surface concentration calculated with \Eq{surface conc
adv-diff eqn}. To solve \Eq{extend surf conc} we use a third-order
Runge-Kutta scheme. The spatial derivatives are calculated by
means of a third-order ENO scheme.

\subsubsection{Bulk concentration field $c$} \label{sec:Numerical model, bulk mass
transport} The evolution of the surfactant concentration field $c$
in the bulk, i.e. outside the droplet, is described by means of
the advection-diffusion equation \beq \pder{c}{t} + \textbf{u}
\cdot \nabla c = D \nabla^2 c \label{Eq:bulk conc adv-diff
eqn}\eeq where $D$ is the diffusivity constant of the surfactant
in the bulk. In order to avoid the introduction of a further
limitation on the time step arising from an explicit treatment of
the diffusive terms, these are treated implicitly. A second-order
Adams-Bashfort scheme is adopted for the time integration \beq
\frac{c^{n+1} - c^n}{\Delta t} = -\frac{3}{2}(\textbf{u} \cdot
\nabla c)^n + \frac{1}{2}(\textbf{u} \cdot \nabla c)^{n-1} + D
\nabla^2 c^{n+1}\label{Eq:bulk conc adv-diff eqn, Adam Bashforth}
\eeq

As the purpose of the present work is the development of a model
that could be adapted to the study of complex microswimmers, where
several chemical species are present dissolved in the bulk and
behave differently \cite{kit11,thu12,ped12}, we implemented
different scenarios, corresponding to different boundary
conditions for $c$ at the droplet's interface $\mathcal{I}$. In
the case that we have described so far, where a surfactant is
dissolved inside the bulk and adsorbed/desorbed at the droplet's
interface, a Neumann boundary condition is required, prescribing
the normal flux $D (\textbf{n} \nabla c)_I$ at the droplet's
interface. This scenario applies, for example, to the surfactant
fields in the system described in Refs.~\cite{thu12, ped12}. Such
a system consists of a liquid crystal droplet, immersed in a
aqueous solution where surfactants are dissolved, both in micellar
and molecular form. A complete treatment of these
mutually-interacting fields is beyond the purpose of the present
work. However, as we wish here to address all the basic mechanisms
at play, we also present a model for the liquid crystal
concentration field diffusing from the droplet into the liquid.
As a consequence of such a diffusion process, the droplet will also shrink.
However, the experiments showed that this process is slow compared to the
droplet's motion and not visible during the time-scales we are interested in
(of the order of seconds and minutes) \cite{ped12}. 
Therefore, the droplet's shrinking will be neglected in our treatment.
The evolution of the liquid crystals dissolved
inside the water can be described by an equation of the same type
of \Eq{bulk conc adv-diff eqn} with a Dirichlet boundary condition
$c_I = c_0$ at the droplet's interface where $c_0$ is a constant
equilibrium value. This model describes the equilibrium between
the pure liquid crystal phase (inside the droplet) and the
dissolved one (outside the droplet). In this case there is of
course no surface concentration field
coupled to the bulk field. \\

- \textit{Neumann boundary condition at the interface}
\\ When the bulk concentration field $c$ and the surface
concentration field interact with each other (e.g. by means of
adsorption/desorption phenomena \cite{thu12, ped12}), we formulate
the boundary condition at the interface as \cite{tei11} \beq D
(\tbf{n} \cdot \nabla c)|_I = j \label{Eq:BC bulk conc} \eeq The
left-hand side represents the net flux of $c$ exchanged at the
interface. It takes positive values when surfactant leaves the
bulk and deposits on the surface. The right-hand side, $j$, is the
increase rate of surfactant on the surface, as derived in
\Eq{coupling term surface/bulk}. Formally, this exchange of
concentration can be incorporated into \Eq{surface conc adv-diff
eqn} by adding a $\delta$-function source term at the interface
\beq \pder{c}{t} + \textbf{u} \cdot \nabla c = D \nabla^2 c -
\delta(\Phi) j \label{Eq:bulk conc adv-diff eqn, thick
interface}\eeq and revising the boundary condition at the
interface to take the form $\tbf{n} \cdot \nabla c|_I = 0$. This
new boundary condition is enforced by fictitiously extending the
field $c$ inside the droplet, and using it to calculate the
time-explicit advection term in \Eq{bulk conc adv-diff eqn, Adam
Bashforth}. To perform this extension, we use the same method
described in Sec.~\ref{sec:Numerical model, surface mass
transport} for the extension of the $\Gamma$ field: we solve to
steady state an equation of the kind of \Eq{extend surf conc},
this time only inside the droplet. We note that the standard
smeared version of delta function \ref{Eq:delta fn} has support on
both sides of the interface. Therefore, in order to guarantee mass
conservation, we need to adopt a new smeared delta function
$\delta_+$, with support only on the outside of the droplet
\cite{zha06}. We define \beq \delta_+(x) =
\begin{cases}
 2\delta_{\epsilon},
& \mbox{ if } 0<x\leq \epsilon \\ 0, & \mbox{ else }
\end{cases} \label{Eq:delta fn, one-sided}\eeq
This procedure has been proved to be rigorous in the
one-dimensional case \cite{zha06}, since the above-defined
$\delta_+$ is a delta sequence function with the same distribution
as the delta function. Unfortunately this property is not exactly
fulfilled for generic higher-dimensional cases, as the numerical
integration over the support does not give the exact value of 1.
However, this approach can still provide good results within the
expected second-order error convergence, as we show in Sec.~
\ref{sec:Validation, adsorption-desorption}. \\

- \textit{Dirichlet boundary condition at the interface} \\
 The implementation of a Dirichlet boundary condition, is carried out
by means of an extrapolated "ghost" bulk concentration field
inside the droplet, to be used for the calculation of the explicit
advective term $(\textbf{u} \cdot \nabla c)^n$ in \Eq{bulk conc
adv-diff eqn, Adam Bashforth}. The adopted procedure is inspired
by the one used in \cite{can12} for the treatment of the
temperature field. The underlying idea is to first calculate the
normal gradient at instant $n$, $c_n^n = (\textbf{n} \cdot \nabla
c)^n$ outside the droplet, by accounting for the Dirichlet
boundary condition, then extrapolate it inside the droplet and
finally use this extrapolated normal gradient field  $c_{n,ext}^n$
to derive an extrapolated bulk concentration field $c_{ext}^n$. In
detail, the procedure works as follows. We compute the physical
normal gradient outside the droplet, by means of a centered
differences second-order discretization. At points adjacent to the
interface, we use a "cutcell method", i.e. we consider a
non-symmetric stencil including the interface. For example, for a
generic point $P$ just at the right of an interface point $I$, the
discrete radial derivative at time instant $n$ will be $
\pder{c}{r}|_P^n = \frac{c_P - c_I}{\theta^n \Delta r}$, where
$\theta^n = (r_P - r_I^n)/\Delta r $. The bulk concentration at
the interface is given by the Dirichlet boundary condition $c_I =
c_0$. After computing the normal gradient outside the droplet, we
extend it inside the droplet, by solving to
steady state the equation \beq \begin{split} \pder{c_{n,ext}}{\tilde{\tau}} - H(-\Phi ^n)(\textbf{n} \cdot \nabla c_{n,ext}) &= 0\\
c_{n,ext}(\textbf{x},0) &= c_n^n(\textbf{x}) \, , \end{split}
\label{Eq:extend bulk conc gradient} \eeq with $\tilde{\tau}$ an
artificial time variable and $H$ the Heaviside function. The
integration of \Eq{extend bulk conc gradient} in the pseudo-time
is performed by means of a second-order Runge-Kutta scheme. The
spatial derivatives of the normal gradient are computed with a
second-order ENO scheme. From the extrapolated normal gradient
$c_{n,ext}$, we derive the values of the extrapolated bulk
concentration field inside the droplet by linear extrapolation, as
$c_{ext} = c_I + \phi (\textbf{n} \cdot c_{n,ext})$.

For the treatment of the time-implicit diffusive terms $D \nabla^2
c^{n+1}$, we also adopt a linear extrapolation of $c^{n+1}$ inside
the droplet, enforcing the boundary condition $c_I = c_0$, when
writing the five-diagonal matrix for the Poisson solver. For
example, for a grid point $P$, characterized by indexes $i,j$ and
laying at the right of the interface point $I$, the discretization
of the Laplacian operator reads \beq \nabla^2 c_{i,j} \simeq
\frac{1}{r_{ij}}\bigg( r_{i+1/2} \frac{c_{i+1,j}-c_{i,j}}{\Delta
r} - r_{i-1/2} \frac{c_{i,j}-c_{i-1,j}^{ext}}{\Delta r} \bigg) +
\frac{c_{i,j+1}-c_{i,j}}{\Delta z} -
\frac{c_{i,j}-c_{i,j-1}}{\Delta z} \label{Eq:discrete Laplacian
operator}\eeq with $c_{i-1,j}^{ext} = c_{i,j} - \frac{c_{i,j} -
c_I}{\theta \Delta r}$. Note that, in \Eq{discrete Laplacian
operator}, we dropped the superscript $n+1$ everywhere for
clarity, but the position of point $I$ used in this
discretization, $r_I$ therefore $\theta$, are those referred to
the new time instant $n+1$, at this point already known from the
solution of \Eq{LS advection eqn}.
\\

\subsection{Chemical reactions}
\label{sec:Numerical model, chemical reactions}

Despite providing some initial motion, in an isolated system,
absorbtion-desorption alone cannot generate a proper
self-sustained propulsion, since they basically only shift
material from one place to another. Unless externally sustained
(like in the case of chemotaxis, that will be discussed in
Sec.~\ref{sec:Results, chemotaxis}) the concentration gradients
are eventually levelled up by diffusion and the droplet stops. In
order to have an actual self-propulsion mechanism, one needs to
consider some additional mechanism, where the energy is actively
used. One possible way is through chemical reactions depleting the
surfactant at the droplet's surface. This happens in droplets
driven by Belousov-Zhabotinsky (BZ) reactions \cite{kit11,thu12}.
We propose hereby a numerical treatment for a simplified scenario,
where the chemical reactions change the nature of the surfactant
itself. We consider two surface concentration fields instead of
one, namely $\Gamma_f$, designating the "fresh" surfactants and
$\Gamma_w$, designating a transformed state of the surfactant, the
"waste". The surfactant in the bulk are in the fresh-state; they
are adsorbed at the surface, then transformed into waste, which
is, in return, desorbed into the bulk. While both fresh and
waste-surfactants contribute to the surface saturation, only the
fresh surfactants contribute to the surface tension. Hence in
Langmuir equation (\ref{Eq:Langmuir eqn}), $\Gamma$ is replaced by
$\Gamma_f$. For each of the two fields we solve an equation of the
kind of \Eq{surface conc adv-diff eqn}, with $j_f$ and $j_w$ (for
the fresh and waste-surfactants respectively) instead of $j$ \beq
j_f = r_a [\Gamma_\infty - (\Gamma_f + \Gamma_w)] c_I - r_c
\Gamma_f \label{Eq:coupling term surface/bulk, fresh} \eeq \beq
j_w = - r_d \Gamma_w + r_c \Gamma_f \label{Eq:coupling term
surface/bulk, waste} \eeq with $r_c$ the constant chemical
conversion rate. We neglect the advection-diffusion process of the
waste-surfactant in the bulk and we take $c$ as the bulk
concentration of fresh surfactant. Hence, instead of \Eq{BC bulk
conc}, the Neumann boundary condition for $c$ at the droplet's
interface will read \beq D (\tbf{n} \cdot \nabla c)|_I = j_f
\label{Eq:BC bulk conc, fresh} \eeq The numerical treatment for
all the concentration fields is the same as outlined in
Sec.~\ref{sec:Numerical model, mass transport}.

\section{Validation}
\label{sec:Validation}
We present here several test cases, separately addressing different aspects of the model.
In the figures, the asterisks denote normalized quantities.
The quantities used for the normalization, denoted by the subscript '0',
are specified in the description of each test case.

\subsection{Surface diffusion}
\label{sec:Validation, surface diffusion}
The present section tackles the surface diffusion of the surfactant (see Sec.~\ref{sec:Numerical model, surface mass
transport}). In particular, we test here the numerical solver (\ref{Eq:surf conc adv-diff eqn, Adam Bashforth})
for the surface diffusion, described in \Eq{surface conc adv-diff
eqn}. To this aim, we considered the case of a static, undeformable spherical
droplet of radius $R_0$, under the assumption of
spherical-symmetry and no exchanges with the bulk ($j = 0$). We
compared the numerical results with the analytical solution of the
surface diffusion equation
\beq \pder{\Gamma}{t}
 = D_s \nabla^2_s \Gamma \label{Eq:surface
conc diff eqn, test}\eeq
with initial condition \beq \Gamma_0(\theta) = \Gamma_0 + \epsilon \sin(\omega \theta)
\label{Eq:Initial surface concentration field} \eeq Here
$\epsilon$ and $\omega$ are two constants and $\theta$ is the angular
position, measured from the $r$ axes in the counterclockwise
direction (see \Fig{Prescribed sigma = sin}). The analytical solution of \Eq{surface conc diff eqn,
test} is \beq \Gamma(\theta, t) = \Gamma_0 + \epsilon
\exp^{-2\omega^2t/R_0^2} \sin(\omega \theta) \label{Eq:Analytic
surface concentration field} \eeq We found an excellent agreement
between the numerical and the analytic results (see \Fig{Surface
conc diffusion TEST}). In \Fig{Convergence, surf conc diffusion
TEST} we display the maximum absolute error of the surface concentration $\Gamma^* = \Gamma / \Gamma_0$ as a
function of time, for different grid sizes: $N_r = 50$, $N_z = 100$
(solid line); $N_r = 100$, $N_z = 200$ (dash-dotted line), $N_r = 200$, $N_z = 400$ (solid line). As time progresses,
a reduction of the grid size $h$ of a factor 2 corresponds to circa a 4-times reduction in the error, as expected from
a second-order method. The time and length scales used for the re-normalization in this test case
are $t_{\Gamma} = R_0^2/D_s$ and $R_0$ respectively .

\begin{figure}
\centering
\includegraphics[width=0.7\textwidth]{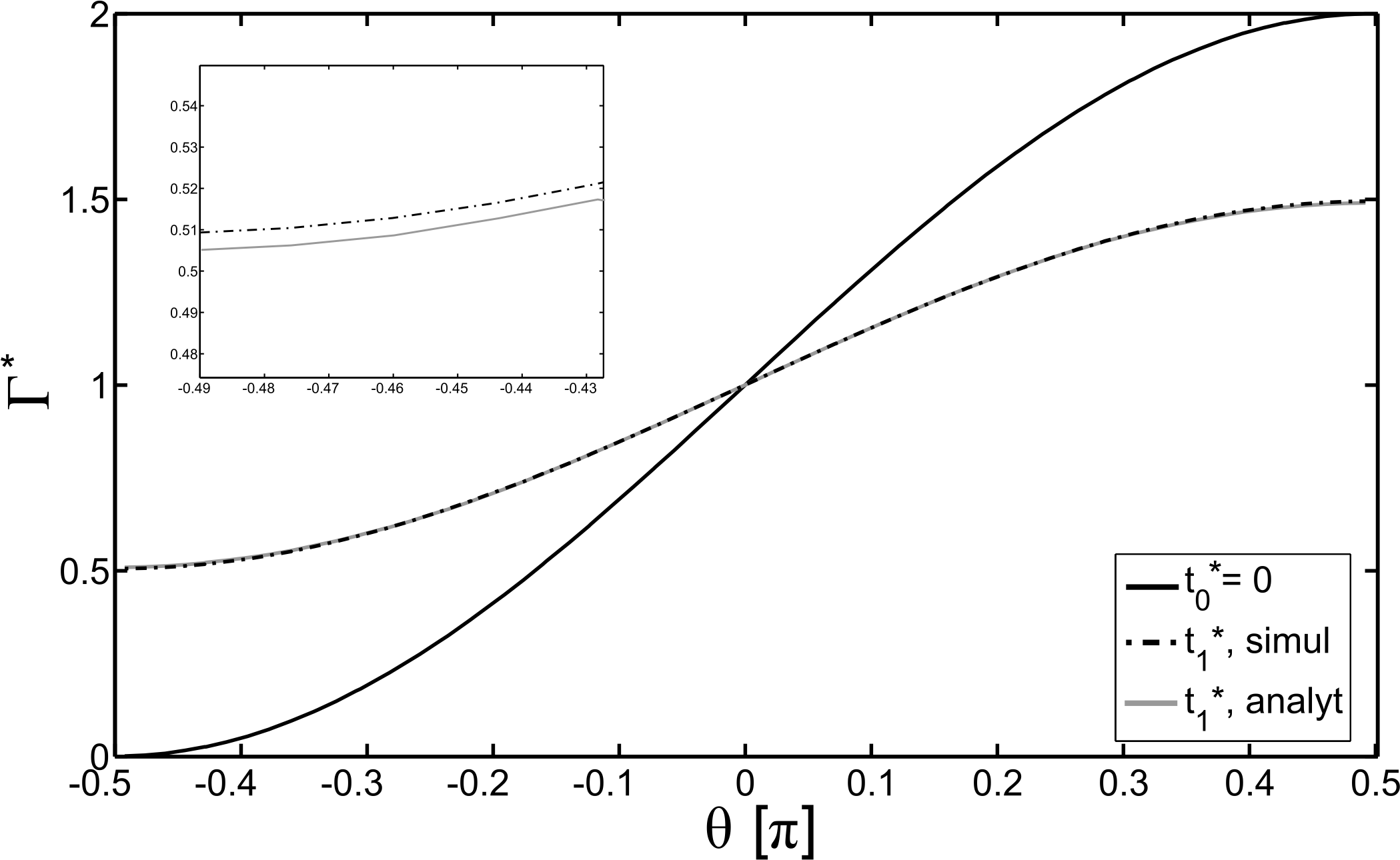}
\caption{Evolution of the surface concentration of a static
undeformable droplet in the spherically-symmetric case, for the
surface diffusion test described in Sec.~\ref{sec:Validation,
surface diffusion}. The black solid line depicts the initial
distribution; after some time, namely at instant $t^* =
0.45$, the numerical results (black
dash-dotted line) show a very good agreement with the analytical
ones (grey solid line). Parameters: $R_0 = 6 \mum$, $\Gamma_0 = 3.4 \cdot 10^{-7} kg/m^2$, $D_s = 10^{-9} m^2/s$
$\epsilon = 1$, $\omega = 1$, $N_r = 100$, $N_z = 200$. }
 \label{Fig:Surface conc diffusion TEST}
\end{figure}

\begin{figure}
\centering
\includegraphics[width=0.7\textwidth]{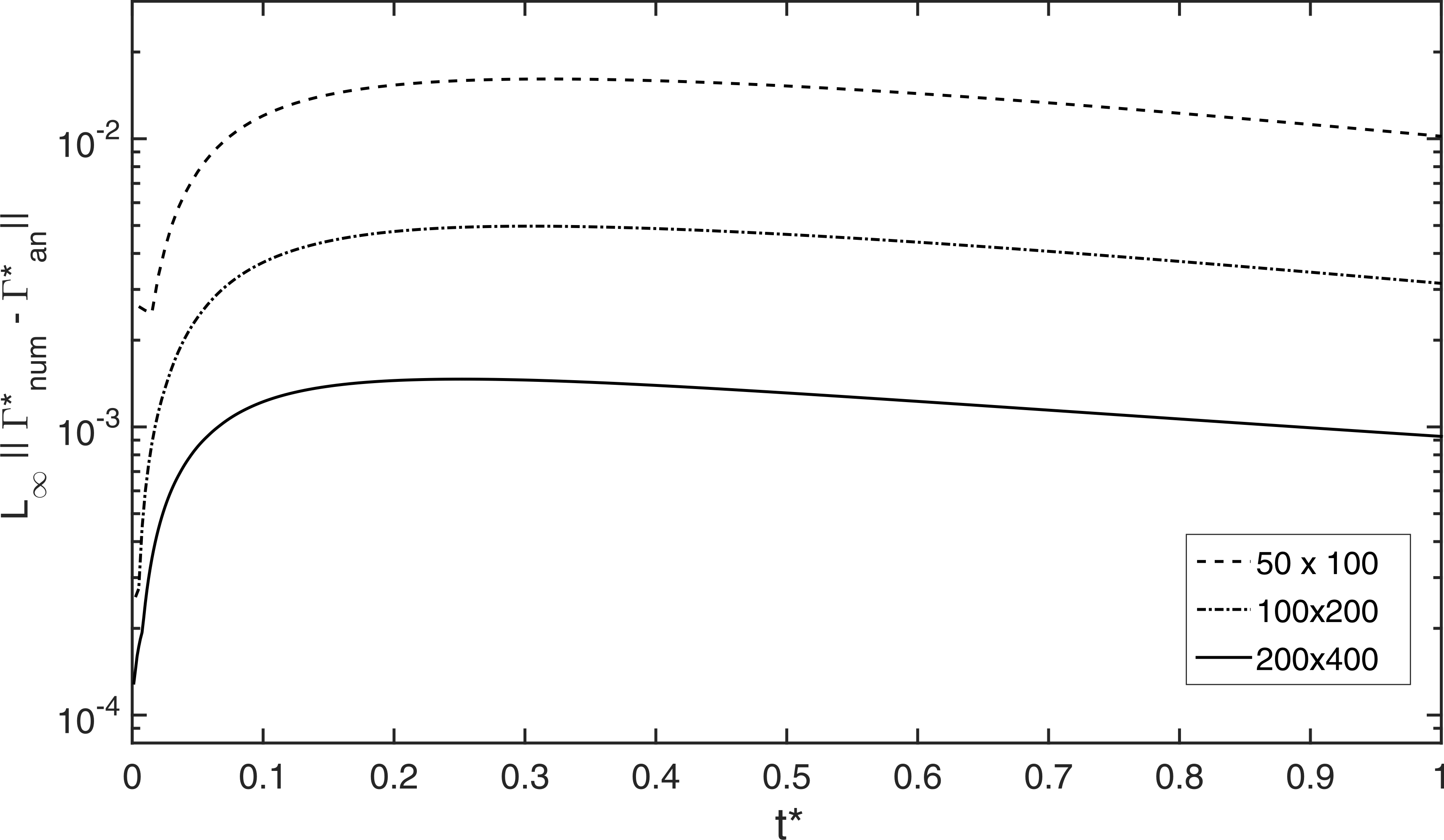}
\caption{Convergence test for surface diffusion: time evolution of
the maximum absolute error $||\Gamma^*_{num} - \Gamma^*_{an} ||$ on
the surface concentration, for a static spherical droplet. Here
the subscript $num$ denotes the numerical results from simulations, $an$ the
analytical ones. Parameters: as in \Fig{Surface conc diffusion TEST}. The two lines
represent different mesh resolutions: $N_r = 50$, $N_z = 100$
(dashed line); $N_r = 100$, $N_z = 200$ (dash-dotted line); $N_r = 200$, $N_z = 400$ (solid line). A
reduction of the grid size $h$ of a factor 2 produces a reduction
of a factor around 4 in the error, as expected from a second-order
method.} \label{Fig:Convergence, surf conc diffusion TEST}
\end{figure}

\subsection{Surface advection and effect of the local change of curvature}
\label{sec:Validation, surface advection}
In the present section we tested the advection of the surface concentration $\Gamma$
and the effects of a local change of curvature at the interface on the field $\Gamma$ (see Sec.~\ref{sec:Numerical model, surface mass
transport}, left hand side of \Eq{surface conc adv-diff eqn}). To
this aim, we considered the case of a deformable, spherical
droplet with initial radius $R_0$ and uniform initial
concentration $\Gamma_0$. We prescribed a normal velocity field
such that $\textbf{u} = \textbf{n} u_n$ and $u_n d =
\textmd{const}$, with $u_n$ the normal component of the velocity
and $d = \sqrt{r^2 + z^2}$ the distance from the center of the
droplet. We neglected the surface diffusion as well as
adsorption/desorption phenomena. With the latter assumption, the
total mass of surfactant on the surface has to be conserved in
time. Hence, the evolution of the field $\Gamma$ is described by
\cite{mur08} \beq \frac{d (\Gamma A)}{dt} = 0 \label{Eq:surface
conc advection eqn}\eeq where $A(t) = 4 \pi R(t)^2$ is the area of
the droplet and $R(t)$ is the radius at time $t$. Upon
integration, one obtains \beq \Gamma(t) = \frac{A_0}{A(t)}
\Gamma_0 \label{Eq:surf conc analytic, expanding droplet}\eeq with
$A_0$, $\Gamma_0$ the initial area and the initial concentration
respectively.
 The analytical results were compared to the numerical ones, after deriving
the average surface concentration at the droplet's surface $\Gamma_{av}$, at each time step as \beq \Gamma_{av} =
\frac{\int_{S_{\textmd{drop}}} \Gamma dS}{S_{\textmd{drop}}}
\label{Eq:Droplet's average surface conc, integral}\eeq with $S_{\textmd{drop}} =
\int_{S_{\textmd{drop}}} dS$ the surface of the droplet. The
integrals have been numerically calculated as \cite{osh03_BOOK}
\beq \Gamma_{av} = \frac{ \int_\mathcal{D} \delta_{\epsilon}(\Phi) \Gamma 2\pi
r dr dz }{ \int_\mathcal{D} \delta_{\epsilon}(\Phi) 2\pi r dr dz}
\label{Eq:Droplet's average surface conc, numerical}\eeq where $\mathcal{D}$
is the computational domain and $\delta_{\epsilon}$ is the smeared
delta function \Eq{delta fn}.
In \Fig{Time-avg surface conc, expanding droplet} we
display the time evolution of the average surface concentration,
derived from simulations (black lines) for different grid sizes: 100x200 (dashed line)
and 200x400 (dot-dashed line) computational cells. The numerical
results are in excellent agreement with the analytical ones
(light-grey solid line). In \Fig{Convergence avg surface conc,
expanding droplet} the time evolution of the relative error on the
average surface concentration is presented. When the grid size $h$
is reduced of a factor 2, the error is reduced of a factor 5,
compatibly with a second-order convergence of the method. The wavelength of the numerical oscillations in
\Fig{Convergence avg surface conc,
expanding droplet} is also reduced of a factor 2, as expected since the time step is

\begin{figure*}
\centering
\includegraphics[width=0.7\textwidth]{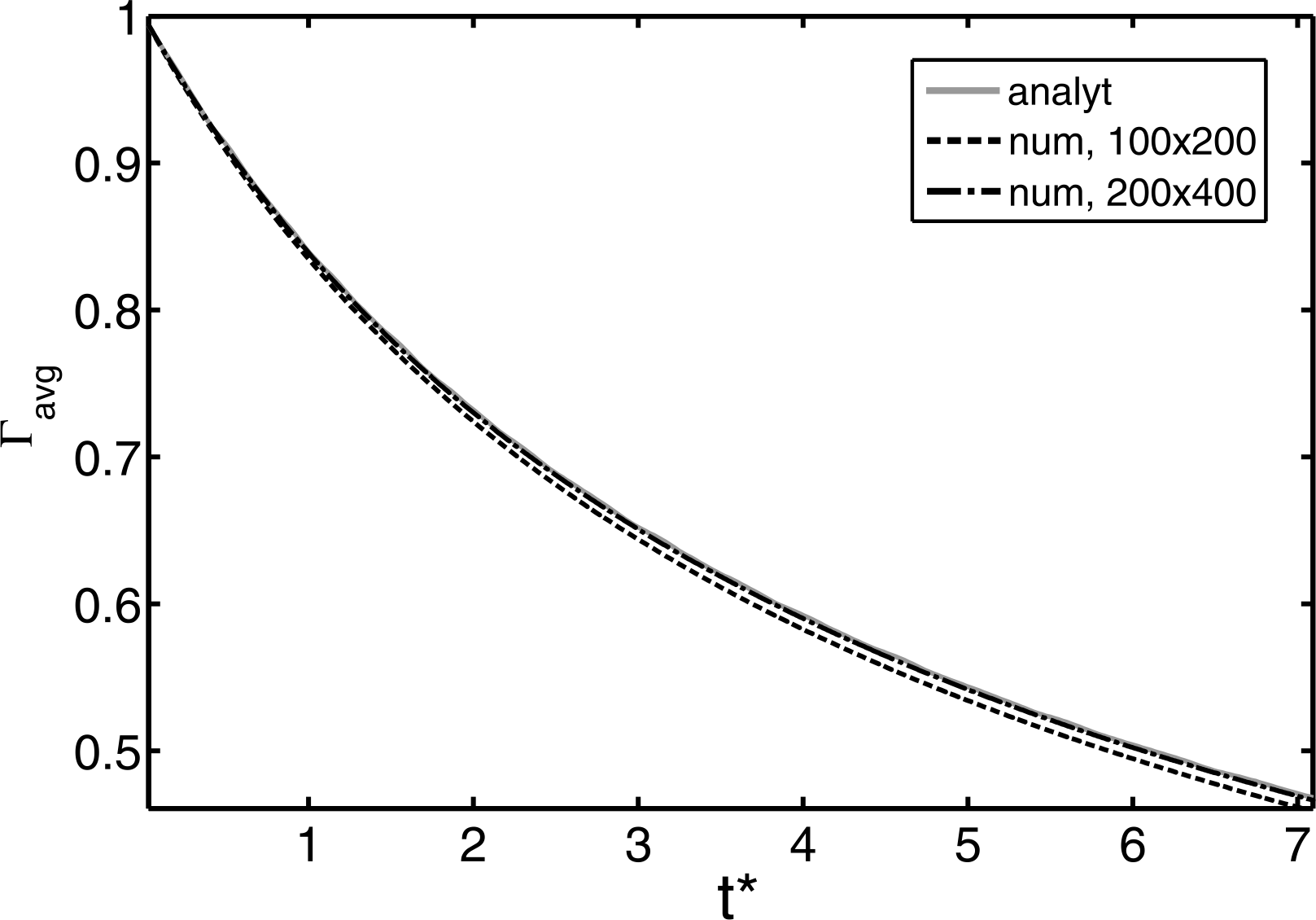}
\caption{Time evolution of the average surface concentration for a
spherical expanding droplet. The numerical results (black lines)
show a very good agreement with the analytical ones (grey solid
line). Parameters: $R_0 = 1 \mum$, 
$\Gamma_0 = 1$. The different black lines represent different mesh
resolutions: $N_r = 100, N_z = 200$ (dashed); $N_r = 200, N_z =
400$ (dash-dotted).}
 \label{Fig:Time-avg surface conc, expanding droplet}
\end{figure*}

\begin{figure*}
\centering
\includegraphics[width=0.7\textwidth]{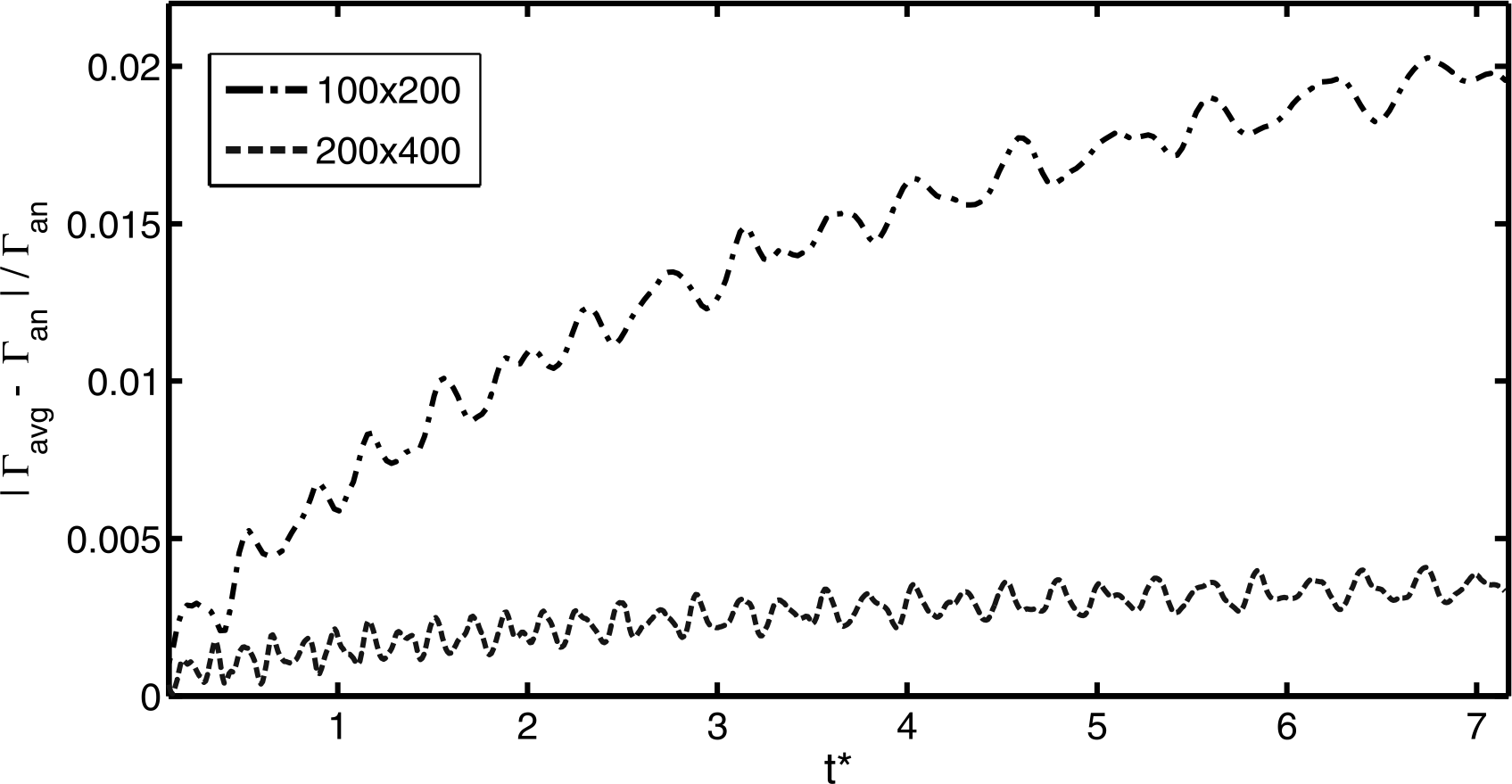}
\caption{Time evolution of the relative error between numerical
and analytical results, for the average surface concentration of a
spherical expanding droplet. Parameters: as in \Fig{Time-avg
surface conc, expanding droplet}. The two lines represent
different mesh resolutions: $N_r = 100, N_z = 200$ (dashed line);
$N_r = 200$, $N_z = 400$ (dash-dotted line) A reduction of the
grid size $h$ of a factor 2 is associated to a reduction of a
factor 5 in the error, compatibly with a second-order convergent
method.}
 \label{Fig:Convergence avg surface conc, expanding droplet}
\end{figure*}

\subsection{Bulk diffusion}
\label{sec:Validation, bulk diffusion}
\begin{figure*}
\centering
\includegraphics[width=0.7\textwidth]{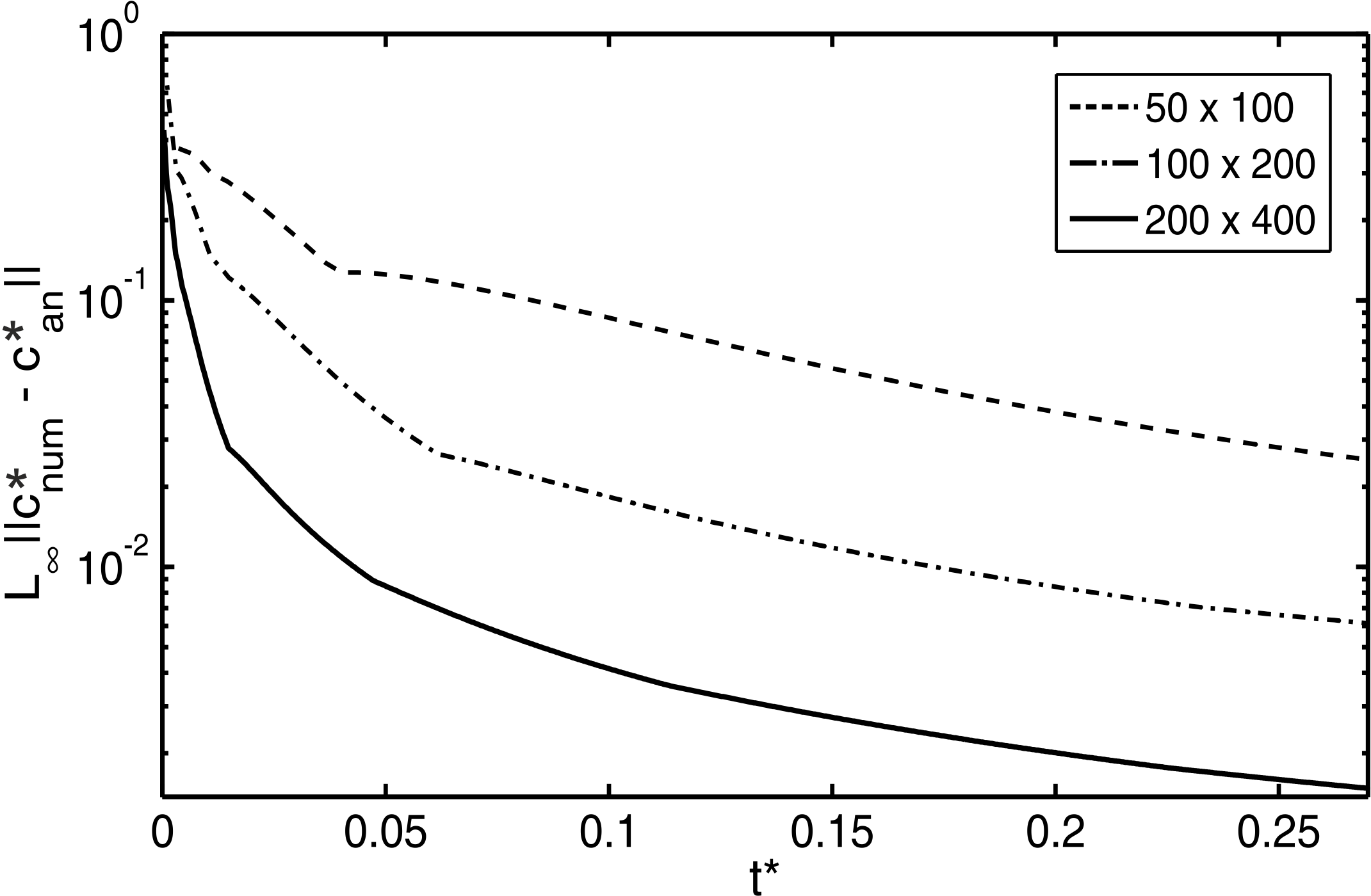}
\caption{Time evolution of the maximum absolute error between the
numerical and the analytical solution, $c_{num}$ and $c_{an}$
respectively, of the bulk diffusion equation, around a static
droplet of radius $R_0$, with initial conditions $c(t = 0) =
c_\infty$ for $r>R_0$ and Dirichlet boundary condition $c(t,r=R_0)
= c_R$ at the interface. Here $t^* = t/t_D$ is the normalized
time,with $t_D = R_0^2/D$ the time scale for diffusion.
Parameters: $R_0 = 25 \mum$, $c_\infty = 1$, $c_R = 0$. The
different lines represent different mesh resolutions: $N_r = 50,
N_z = 100$ (dashed); $N_r = 100, N_z = 200$ (dash-dotted); $N_r =
200, N_z = 400$ (solid). After an initial transient, a reduction
of the grid size $h$ of a factor 2 is associated to a reduction of
a factor 4.5 in the error, as expected from a second-order
method.}
 \label{Fig:Bulk conc convergence}
\end{figure*}

In the present section we address the diffusion of the bulk concentration field $c$ (see Sec.~\ref{sec:Numerical model, bulk mass
transport}).
In particular, we test the
numerical solver (\ref{Eq:bulk conc adv-diff eqn, Adam Bashforth})
for the diffusion of the bulk concentration, described in \Eq{bulk
conc adv-diff eqn}, with Dirichlet
boundary condition at the droplet's interface. To this aim we derived the analytical solution of the
diffusion equation \beq \pder{c}{t}
 = D \nabla^2 c \label{Eq:bulk
conc diff eqn, test}\eeq on the outside of a spherical
undeformable static sphere of radius $R_0$ in the spherically
symmetric case. We took as boundary conditions $c = c_\infty$ in
the far field (ideally $r \rightarrow \infty$) and Dirichlet
boundary condition $c(t,R_0) = c_R$ at the droplet's interface.
The initial condition was chosen as $c(t = 0) = c_\infty$ for
$r>R$. The analytic solution of such a problem is \beq
c_{an}(r,t)= \frac{R_0}{r} (c_R - c_\infty)
\textrm{erfc}\biggl(\frac{r-R_0}{2 \sqrt{Dt}}\biggr) + c_\infty
\label{Eq:Analytic bulk concentration}\eeq We found a very good
agreement between analytical and numerical results (see \Fig{Bulk
conc analytic-num}). In \Fig{Bulk conc convergence} we show the
maximum absolute value of the error between numerical and
analytical results, respectively $c_{num}$ and $c_{an}$, as a
function of time, for different grid sizes. After an initial
transient, reducing the grid size $h$ of a factor 2 implies a
reduction of a factor 4.5 in the error, as expected from a
second-order method.

\begin{figure*}
\[
\raisebox{0.4\textwidth}{\text{(a) \ }}
\includegraphics[height=0.4\textwidth]{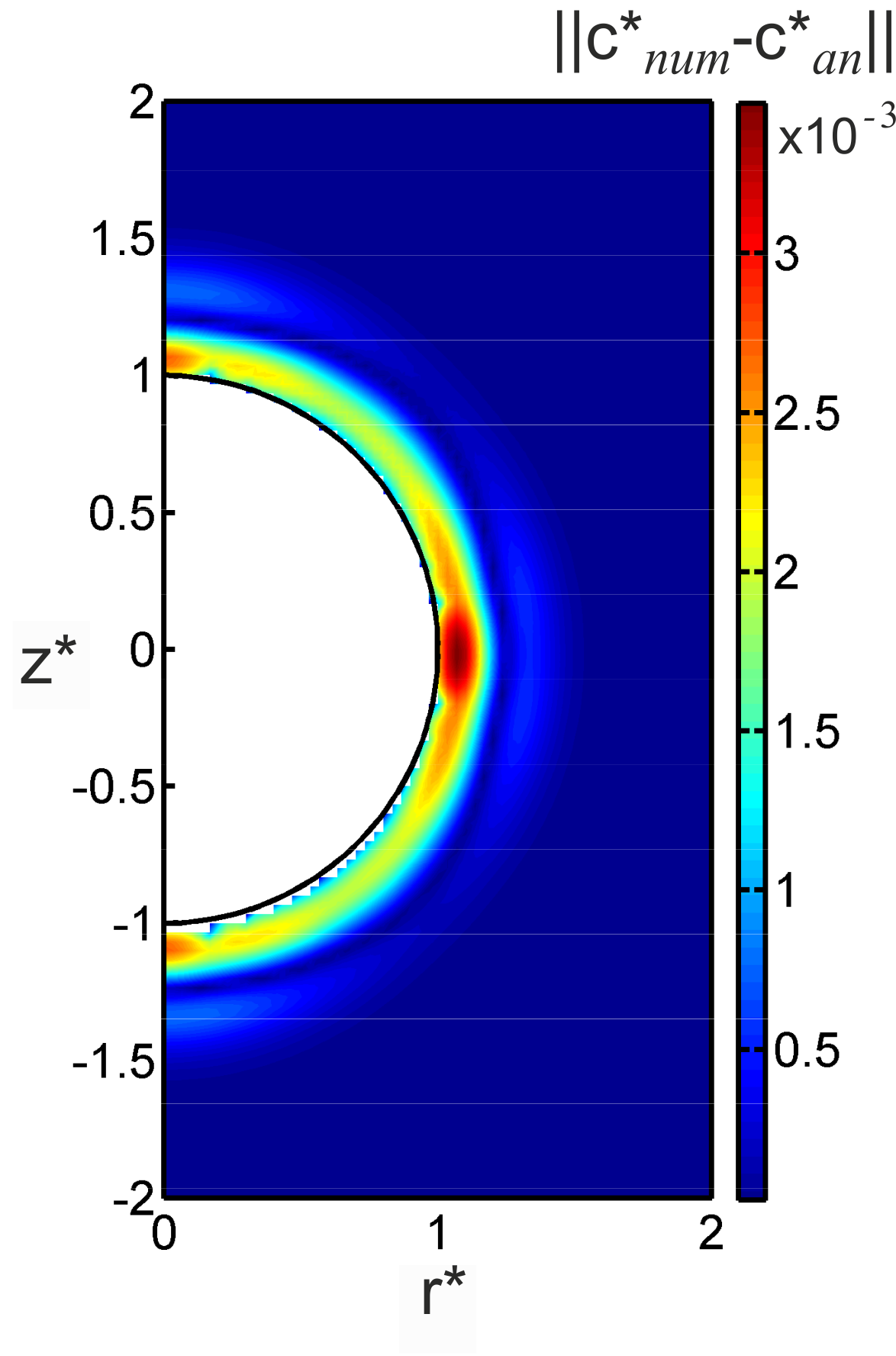}
\qquad\qquad \raisebox{0.4\textwidth}{\text{(b) \ }}
\includegraphics[height=0.4\textwidth]{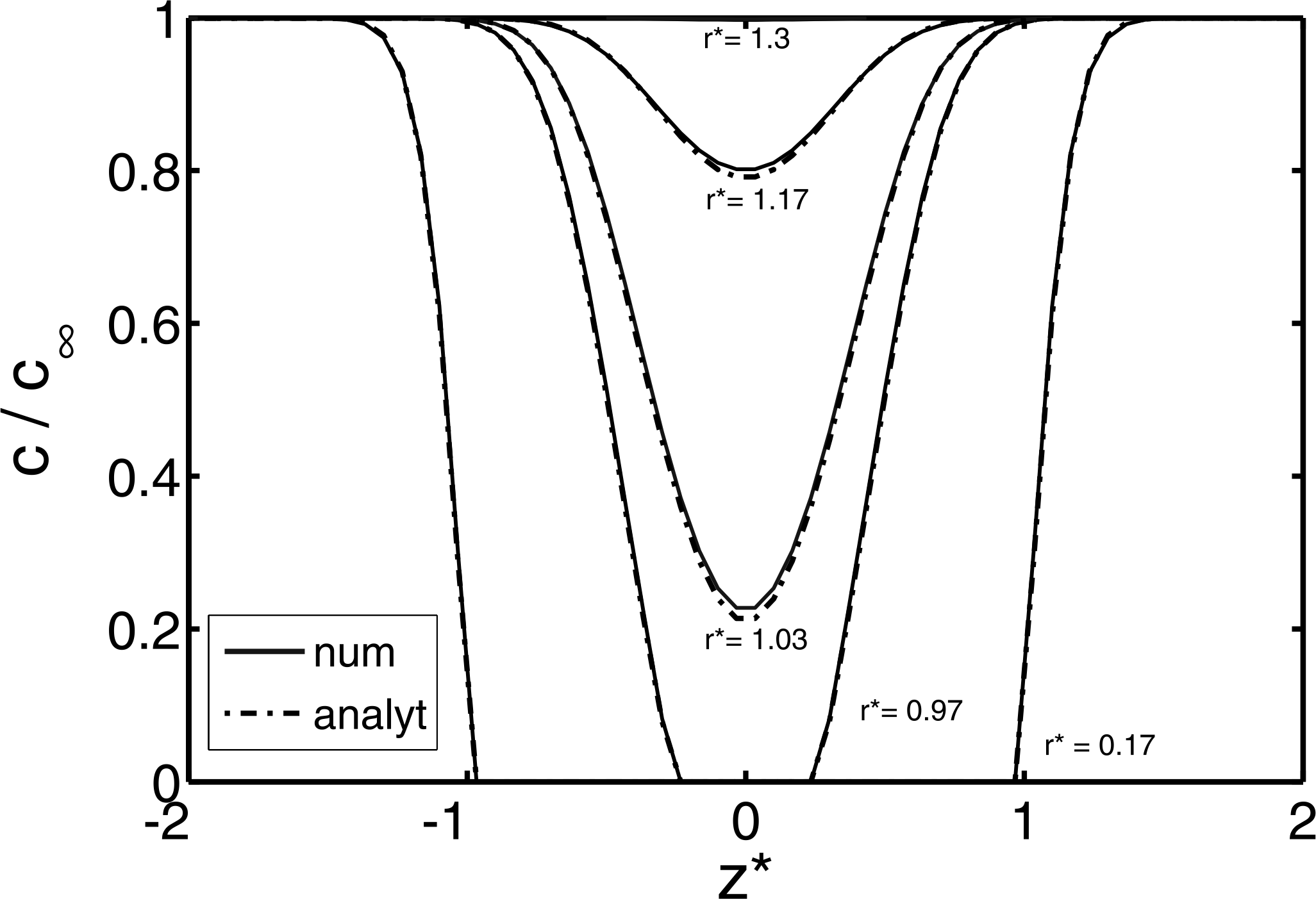}
\]
\caption{Absolute error on $c^*=c/c_\infty$ between the analytical and
the numerical solution for the diffusion equation in spherical
coordinates, around a non-moving droplet of radius $R$, with
initial condition $c(t = 0,r)= c_\infty H(\Phi)$ and Dirichlet
boundary condition $c(r=R_0) = c_R$ at the interface. Parameters:
$R_0 = 25 \mum$, $c_\infty = 1$, $c_R = 0$,  $N_r = 100$, $N_z =
200$ (a) absolute error at each point of the computational domain,
at instant $t/t_D = 0.435$ (with $t_D = R_0^2/D$ the time scale
for diffusion) (b) section of the numerical (solid lines) and
analytical (dash-dotted lines) values of the bulk concentration,
at time $t/t_D = 0.435$, along the z-axes for different values of
$r^*=r/R$.}
 \label{Fig:Bulk conc analytic-num}
\end{figure*}

\subsection{Adsorption-desorption}
\label{sec:Validation, adsorption-desorption}

In the present section we test the solution of the bulk concentration diffusion equation, subject to Neuman
boundary condition at the droplet's interface (see Sec.~\ref{sec:Numerical model, bulk mass
transport}). We tackle the coupling between the surface concentration field $\Gamma$ and the bulk concentration field $c$. In particular
we address the adsorption-desorption mechanism described by Eqs.~(\ref{Eq:coupling
term surface/bulk}) and (\ref{Eq:BC bulk conc}) and its numerical implementation \Eq{bulk conc adv-diff eqn, thick
interface}.

\subsubsection{Test 1. Residual method}
\label{sec:Validation, residuals adsorption}
We test hereby the numerical implementation (\ref{Eq:bulk conc adv-diff eqn, thick
interface}) of the boundary condition (\ref{Eq:BC bulk conc}), thus showing that
the adopted numerical procedure is appropriate to address the Neumann boundary condition
on the bulk concentration field $c$ at the droplet's interface. To this aim, we
adopted the following procedure. We neglected the advection of the
liquid ($\textbf{u} =0$) and we considered the bulk diffusion
equation, \Eq{bulk conc diff eqn, test}, in spherical coordinates
in the case of a static, undeformable droplet of radius $R_0$. In
a general case, if a functional form for the concentration field
is prescribed, $c^{\textmd{test}}(t,\textbf{x})$, \Eq{bulk conc
adv-diff eqn} will not be satisfied exactly, but a residual
$f(t,\textbf{x})$ will be present, such that \beq
\pder{c^{\textmd{test}}}{t}= D \nabla^2 c^{\textmd{\textmd{test}}}
+ f(t,\textbf{x}) \label{Eq:Test absorbt/desorpt, residual} \eeq
Similarly, the boundary condition at the droplet interface will be
\beq \textbf{n}\cdot \nabla c^{\textmd{test}} |_I =
g(t,\textbf{x}_I) \label{Eq:Test absorbt/desorpt, BC normal flux}
\eeq with $g(t,\textbf{x}_I)$ a known function. If one solves
numerically \Eq{Test absorbt/desorpt, residual}, with boundary
condition \Eq{Test absorbt/desorpt, BC normal flux} at the
interface and initial condition
$c^{\textmd{test}}({t=0,\textbf{x}})$, the prescribed
$c^{\textmd{test}}$ should be retrieved, instant by instant, a
part from the error introduced by the numerical method. Measuring
this error and assessing its order of convergence will therefore
indicate if the numerical method is really solving the equation
that it is supposed to solve.

In our test case, we chose a function $c^{\textmd{test}}$
satisfying the boundary conditions on the borders of the
computational domain $\mathcal{D}$, i.e. such that
$c^{\textmd{test}}|_{\partial \mathcal{D}} = 0$.
In particular, we took the test function
 \beq
c^{\textmd{test}}(t,\textbf{x}) =
\begin{cases}
 \textmd{e}^{-t} y \textmd{e}^{-\frac{1}{1-y}}, & \mbox{ if } 0< y < 1 \\
  0, & \mbox{ else }
\end{cases}
\label{Eq:Test absorpt/desorpt, prescribed fn} \eeq where $y =
(r^2+z^2 - R_0^2)/(L^2 - R_0^2)$, with $R_0$ the radius of the
droplet and $L=2R_0$. We substituted it inside \Eq{Test
absorbt/desorpt, residual} and \Eq{Test absorbt/desorpt, BC normal
flux}, thus analytically deriving $f(t,\textbf{x})$ and
$g(t,\textbf{x})$. By applying the numerical thick interface
method described in \ref{sec:Numerical model, mass transport}
to solve \Eq{Test absorbt/desorpt, residual} with Neumann boundary
condition \Eq{Test absorbt/desorpt, BC normal flux}, we
then derived the numerical solution and we calculated the maximum
absolute error respect to the exact one $c^{test}$. In
\Fig{Adsorption TEST, residuals} we show a comparison between
numerical and analytical results of the concentration distribution
along the first column of the computational domain (at $r = h/2$),
for different time instants. The two provide good agreement. In
\Fig{Residuals adsorption TEST, convergence} we plot the time
evolution of such an error, for different grid sizes: $N_r = 100$,
$N_z = 200$ (dash-dotted line), $N_r = 200$, $N_z = 400$ (solid
line). After an initial transient, a reduction of the grid size
$h$ of a factor 2 implies a reduction of a factor 4 in the error,
in agreement with a second order convergence of the method.

\begin{figure}
\centering
\includegraphics[width=0.7\textwidth]{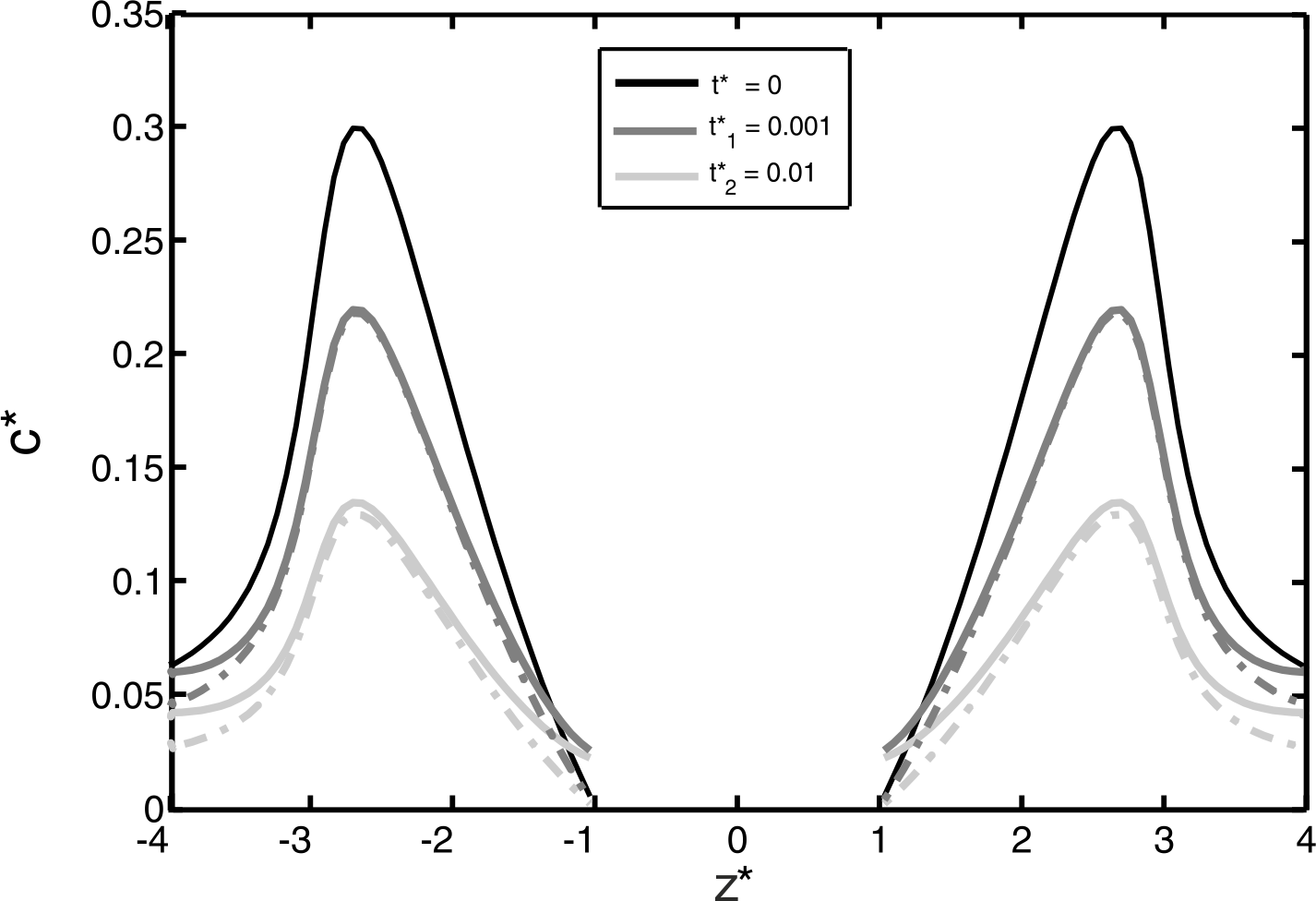}
\caption{Comparison of numerical (solid lines) and analytical
(dash-dotted lines) results of residuals test for surfactants
adsorption, as outlined in Sec.~\ref{sec:Validation, residuals
adsorption}. The figure represents the bulk concentration
distribution along the first column of the computational domain
for different time instants: $t^* = 0$ (black), $t^*_1 = 0.001$
(dark-grey) and $t^*_2 = 0.01$ (light-grey). The axial coordinate
is expressed as $z^*=z/R_0$. Parameters: $R = 6 \mum$, $N_r = 100$,
$N_z = 200$. }
 \label{Fig:Adsorption TEST, residuals}
\end{figure}

\begin{figure}
\centering
\includegraphics[width=0.7\textwidth]{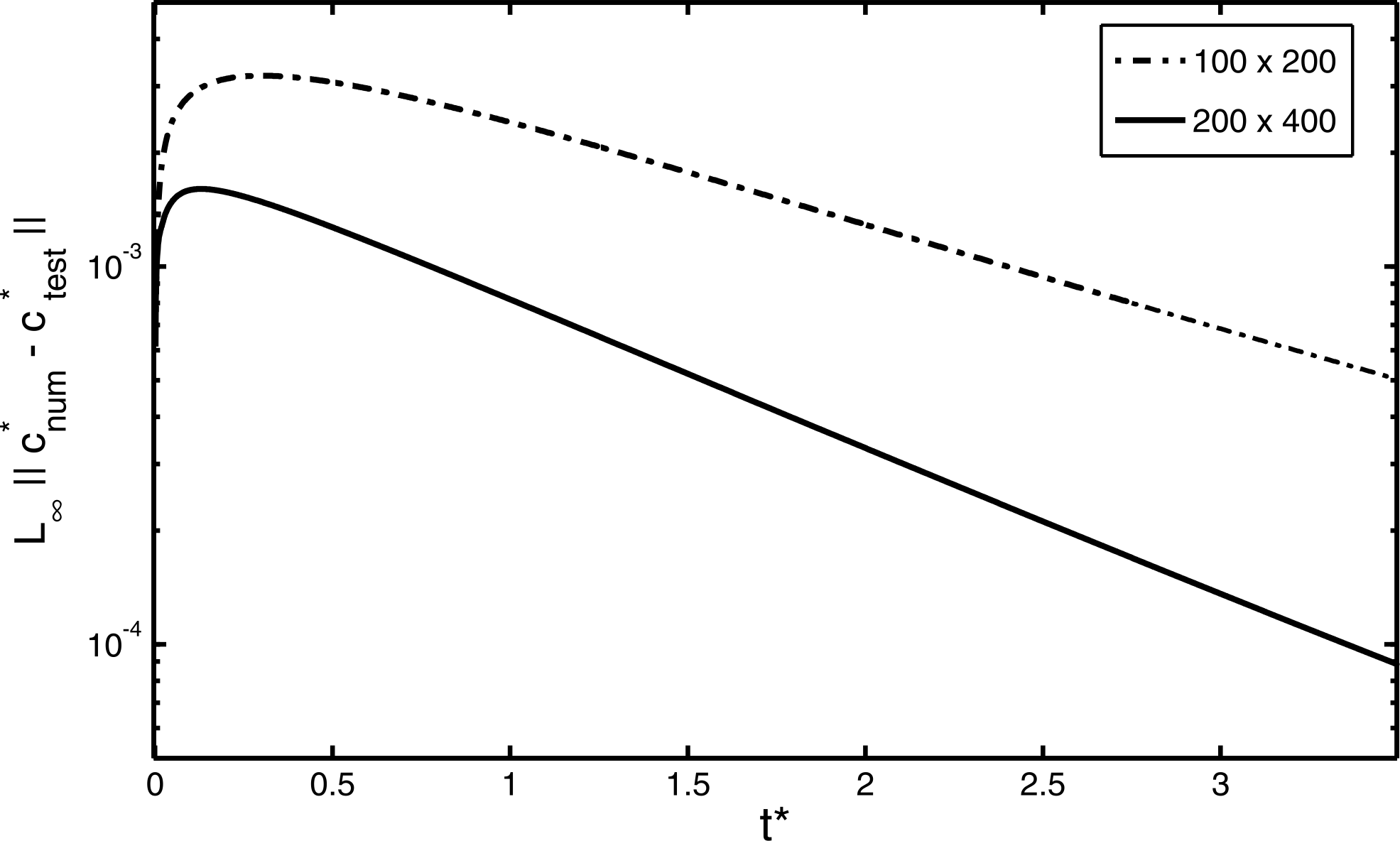}
\caption{Convergence test for residual test for surfactant
adsorption, as outlined in Sec.~\ref{sec:Validation, residuals
adsorption}: time evolution of the maximum absolute error on the
bulk concentration, for a static undeformable spherical droplet of
radius $R_0$. Parameters: $R_0 = 6 \mum$. The different lines
represent different mesh resolutions: $N_r = 100$, $N_z = 200$
(dash-dotted line), $N_r = 200$, $N_z = 400$ (solid line). A
2-times reduction of the grid size $h$ corresponds to a 4-times
reduction in the error, as expected from a second-order convergent
method.} \label{Fig:Residuals adsorption TEST, convergence}
\end{figure}

\subsubsection{Test 2. Analytical solution (Muradoglu \textit{et al.}, J.Comput.Phys., 2008)}
\label{sec:Validation, Muradoglu adsorption}
In the present test case, we address the solution of the bulk concentration diffusion equation,
subject to a simplified Neumann boundary condition, where only adsorption occurs from the
bulk to the surface and no desorption (\Eq{BC bulk conc}, with $r_d = 0$).
To this aim, we considered the case of a static undeformable spherical droplet of
radius $R$, where only bulk diffusion and adsorption were present.
Under these assumptions, the equation to be solved became \beq
\pder{c}{t} = \frac{D}{r^2} \pder{}{r}\bigg( r^2 \pder{c}{r}
\bigg) \label{Eq:Adsorption test (Muradoglu), bulk conc eqn}\eeq
with boundary condition at the interface $D \textbf{n} \cdot
\nabla c|_I = j$ and $j = r_a c_I$; therefore \beq \frac{d
\Gamma}{dt} = r_a c_I \label{Eq:Adsorption test (Muradoglu), surf
conc eqn} \eeq The bubble had an uniform initial surface
concentration $\Gamma_0 = 0$. The initial bulk concentration was
also uniform and equal to $c_0$. An analytical solution of this
set of equations has been derived for the case of an infinite
domain in \cite{mur08,mur14} \beq \frac{c_0 - c}{c_0} = \frac{r_a
\sqrt{\pi D t} }{D \bigg[1 + \frac{\sqrt{\pi D t}}{R}\bigg( 1 +
\frac{r_a R}{D} \bigg) \bigg]} \frac{R}{r} \textrm{erfc} \bigg(
\frac{r-R}{2\sqrt{D t}} \bigg) \label{Eq:Adsorption test
(Muradoglu), bulk conc analytic} \eeq It has been pointed out in
\cite{mur14}, that such a solution is mostly valid for small
times. The characteristic scales to use for the
nondimensionalization of the equations in this test case are $L_0
= R_0$ for lengths and $T_0 = R_0^2/D$ for time. We found
a good agreement between numerical and analytical results
\Fig{Adsorbtion TEST, Muradoglu}.

\begin{figure}
\centering
\includegraphics[width=0.7\textwidth]{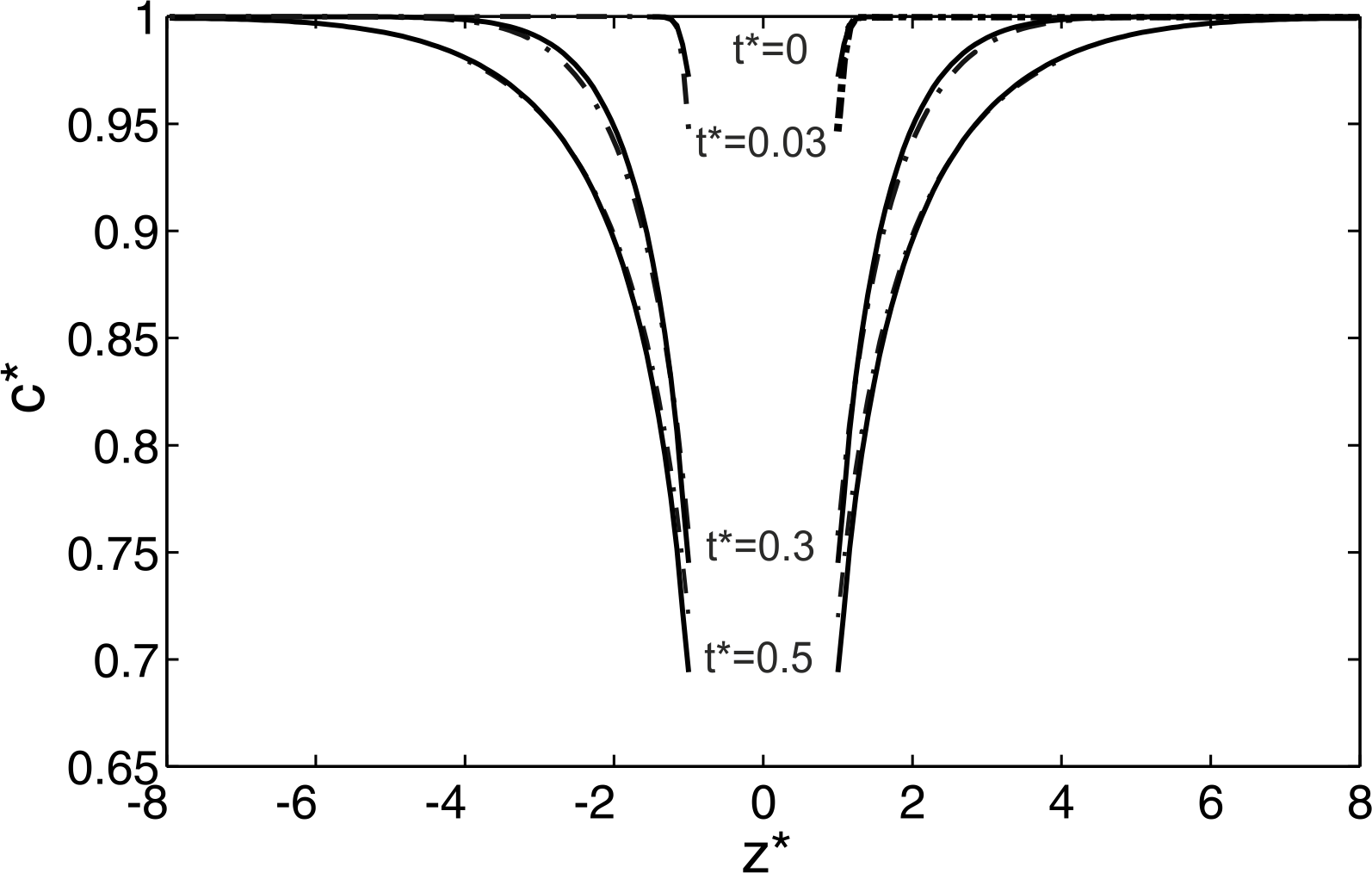}
\caption{Comparison of numerical and analytical results for the
test for surfactant adsorption as outlined in
Sec.~\ref{sec:Validation, Muradoglu adsorption}. The figure
represents the bulk concentration distribution along the first
column of the computational domain for different time instants.
The axial coordinate is expressed as $z^* = z/R_0$. Parameters:
$R_0 = 6 \mum$, $N_r = 150$, $N_z = 300$. }
 \label{Fig:Adsorbtion TEST, Muradoglu}
\end{figure}


\subsection{Marangoni test}
\label{sec:Validation, Marangoni test}
\begin{figure}
\centering
\includegraphics[width=1.\textwidth]{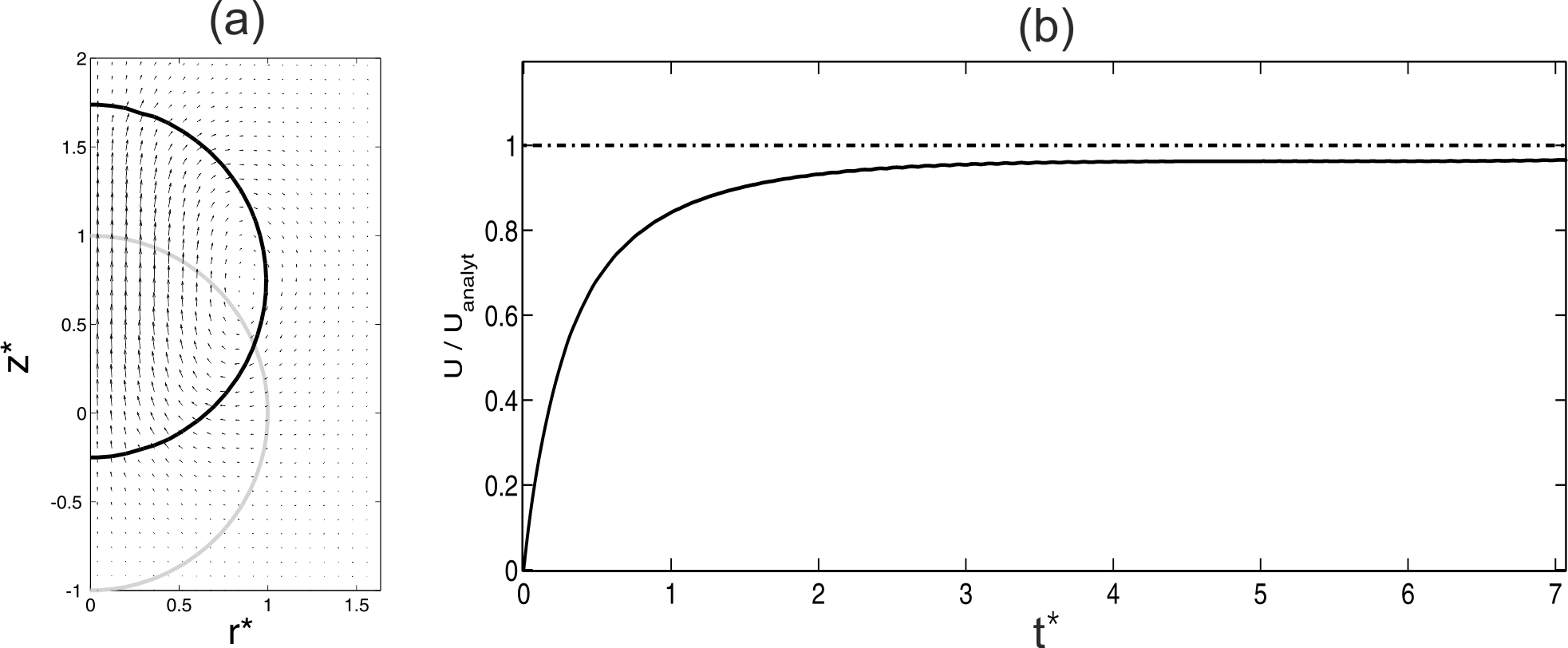}
\caption{Marangoni test for a spherical droplet immersed in a
concentration gradient, as outlined in \ref{sec:Validation,
Marangoni test}. Parameters: $R_0 = 6 \mum$ , $\beta = 1$,
$\sigma_0 = \Gamma_0 = 1$, $N_r = 100$, $N_z = 200$. (a) Velocity
field at time $t_1^*=6.04$ in the static frame of reference
(arrows). The grey line represents the position of the droplet at
time $t^* = 0$, the black line at $t^* = t_1^*$. (b) Time
evolution of the average velocity from numerical simulations
(solid line), in comparison to the analytical terminal velocity
derived by Young et al. \cite{you59} (dash-dotted line). The
difference lies within the $4 \%$.}
 \label{Fig:Marangoni test}
\end{figure}

In the present test, we address the flow solver described in Sec.\ref{sec:Numerical model, dynamics}
as well as the numerical treatment of the kinematics \ref{sec:Numerical model, kinematics} of
droplets moving by Marangoni flow, due to the presence of a non-uniform surface tension on the interface.
To this aim, we consider a spherical
undeformable droplet of radius $R_0$, inside a cylindrical channel
of length $L$. We impose a linear variation of the surface
concentration field $\Gamma$, through the length of the channel,
such that \beq \Gamma(z) = \Gamma_{\infty} \frac{z}{L}
\label{Eq:Surf conc field, Marangoni test}\eeq where
$\Gamma_{\infty}$ is a constant. The surface tension obeys the law
\beq \sigma(z) = \sigma_0 \bigg(1 - \beta
\frac{\Gamma(z)}{\Gamma_{\infty}} \bigg) \label{Eq:Surface tension
field, Marangoni test}\eeq in which $\sigma_0$ is the surface
tension of a clean surface, $\mu$ is the viscosity of the fluid
both outside and inside the droplet and $\beta$ is a constant. The
droplet is therefore expected to move along the concentration
gradient, solely due to Marangoni stresses. For such a case, an
analytical solution for the terminal velocity has been derived,
under the approximation of Stokes flow \cite{you59, mur08} \beq
U_{an} = \frac{2 \beta \sigma_0 R_0}{ 15 L \mu} \label{Eq:Terminal
velocity, Marangoni test}\eeq In the simulation, we derived the
value of the extrapolated field $\Gamma$ at the center of each
computational cell by first finding the closest point $I$ of the
interface, by means of \Eq{closest interface point} and using
\Eq{Surf conc field, Marangoni test} with $z = z_I$. The velocity
of the droplet at each time step was computed as \beq U =
\frac{\int_{V_{\textmd{drop}}} u_z dV}{V_{\textmd{drop}}}
\label{Eq:Droplet's velocity, integral}\eeq with $u_z$ the
vertical component of the velocity and $V_{\textmd{drop}} =
\int_{V_{\textmd{drop}}} dV$ the volume of the droplet. The
integrals have been numerically calculated as \cite{osh03_BOOK}
\beq U = \frac{ \int_\mathcal{D} [1 - H_{\epsilon}(\Phi)] u_z 2\pi
r dr dz }{ \int_\mathcal{D} [1 - H_{\epsilon}(\Phi)] 2\pi r dr dz}
\label{Eq:Droplet's velocity, numerical}\eeq where $\mathcal{D}$
is the computational domain and $H_{\epsilon}$ is a smeared
Heaviside function \cite{xu12} \beq H_{\epsilon}(x) =
\begin{cases} 0, & \mbox{ if }x < -\epsilon \\
 \frac{1}{2} \left( 1 + \frac{x}{\epsilon} + \frac{1}{\pi} \textmd{sin} \frac{\pi x}{\epsilon} \right),
& \mbox{ if }|x|\leq \epsilon \\
1, & \mbox{ if }x > \epsilon
\end{cases} \label{Eq:Heaviside fn}\eeq
The appropriate characteristic scales to use for the
non-dimensionalization of the equations in this test case are $L_0
= R_0$ for the lengths and $T_0 = \sqrt{\frac{2\pi \rho R_0^2 L
}{\beta \sigma_0}}$ for the time, corresponding to a displacement
of around one radius length.
 In \Fig{Marangoni test}.a the grey line is the
initial contour of the droplet, the black one is the contour at
time $t_1^*=6.04$. The black arrows represent the velocities at
time $t_1^*$. In \Fig{Marangoni test}.b we show the time evolution
of the average velocity of the droplet as derived from simulations
(solid line), and we compare it with the analytical solution. With
a computational grid of 100x200 cells, when the steady-state is
reached, we find agreement within $3.5 \%$.

\subsection{Mass conservation}
\label{sec:Validation, mass conservation} 
\begin{figure}
\centering
\includegraphics[width=0.7\textwidth]{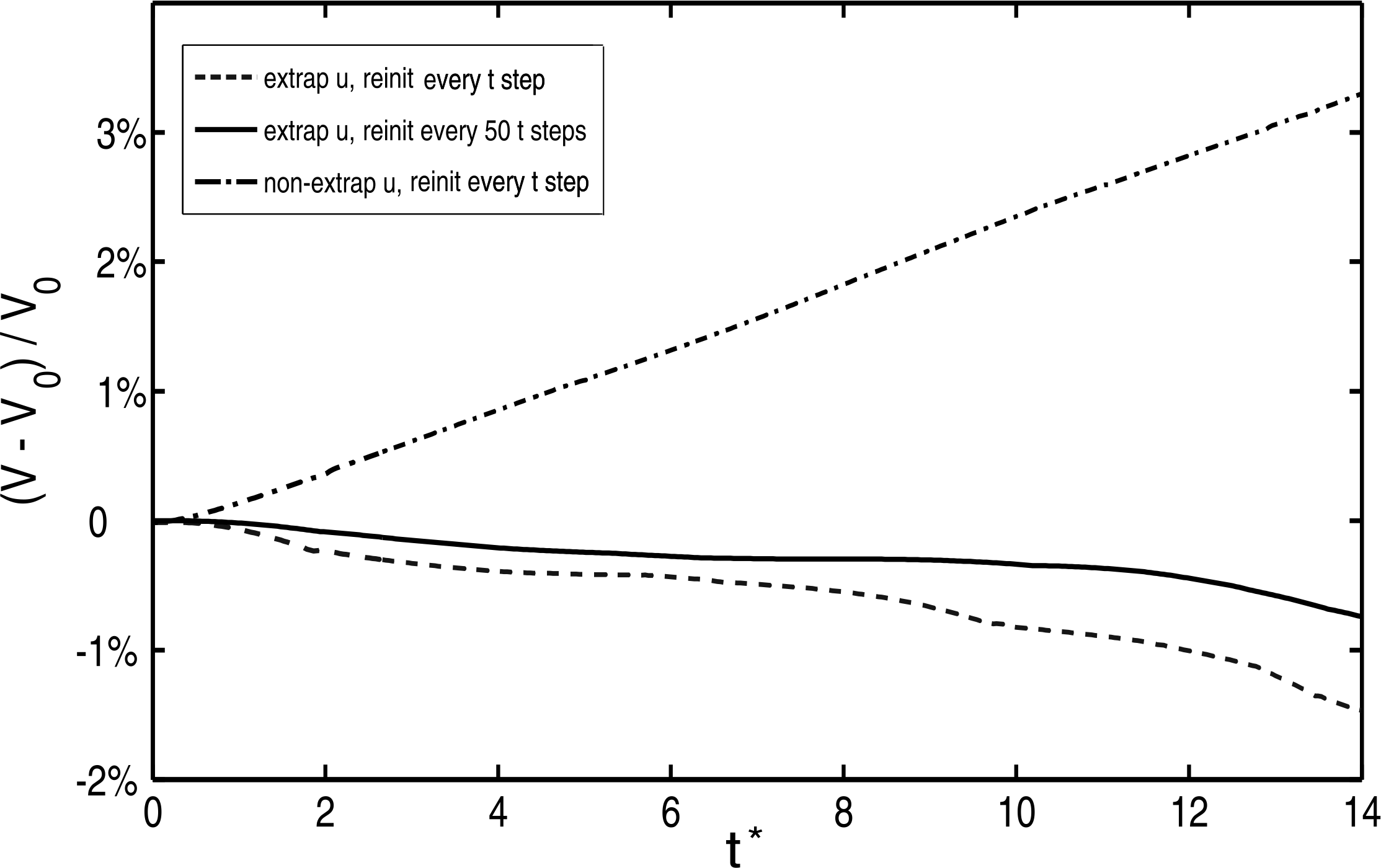}
\caption{Time evolution of the relative error on the droplet
volume, representing the mass loss during the simulation of a
droplet moving by Marangoni flow, described in
Sec.~\ref{sec:Validation, Marangoni test}. We show the results for
different methods: using the real velocity field $\textbf{u}$ to
advect the interface, reinitializing at each time step
(dash-dotted line); using the extended velocity field
$\textbf{u}_{ext}$, reinitialization every time step (dashed-line)
or every 50 time steps (solid line). In the last case the mass
loss error is below $1\%$ for a displacement of the droplet of
around 10 radii. }
 \label{Fig:Mass loss}
\end{figure}

The level set method is
known to suffer from artificial mass loss, mainly originating from
the reinitialization step. In our treatment we address this
problem, by using an extrapolated velocity field (built in the way
described in Sec.~\ref{sec:Numerical model, kinematics}) to advect
the level set function, \Eq{LS advection eqn} and reducing the
frequency of reinitialization steps. Thus we perform a
reinitialization only when required (see Sec.~\ref{sec:Numerical
model, kinematics}). In this test, we wanted to assess the
beneficial nature of this procedure. To this aim, we adopted the
same setting of Sec.~\ref{sec:Validation, Marangoni test}, a
droplet moving due to Marangoni stresses, in response to a
gradient in the surface concentration field. We examined the mass
loss by tracking the change of volume, since the fluid density is
constant in our setup. In \Fig{Mass loss} we display the time
evolution of the relative error on the volume $V$, respect to the
initial volume $V_0$, for different numerical approaches: using
the real velocity field $\textbf{u}$ to advect the interface and
reinitializing at each time step (dash-dotted line); using the
extended velocity field $\textbf{u}_{ext}$, reinitialization every
time step (dashed-line) or every 50 time steps (solid line). We
show that, with the latter technique, it is possible to reduce the
mass loss below $1\%$ for a displacement of the droplet of around
10 radii, corresponding to several thousands of computational
steps.

\section{Propulsion mechanisms}

In the following section, we present several scenarios that may occur in
the study of active droplets driven by Marangoni flow.

\subsection{Pusher/puller behavior of "squirmers"}
\label{sec:Results, pusher/puller behaviour}
\begin{figure} [t]
\centering
\includegraphics[width=0.32\textwidth]{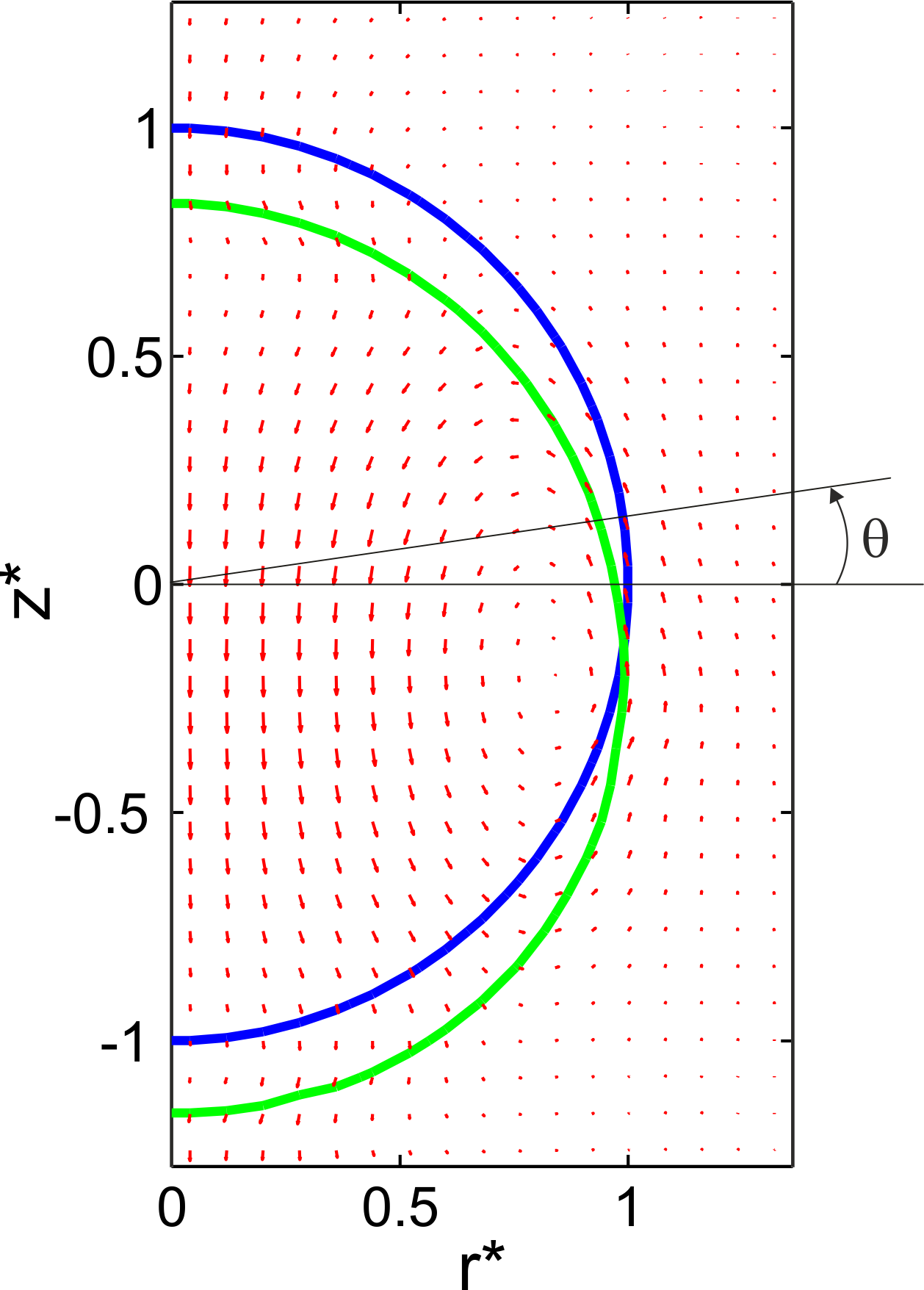}
\caption{Numerical simulation of a droplet moving downwards by
Marangoni flow, when a surface tension field is prescribed as
$\sigma(\theta) = \sigma_0 (1 + \epsilon \sin\theta/2)$.
Parameters: $R_0 = 25 \mum$, $\epsilon = 0.5$, $N_r = 50$,  $N_z = 100$. The black line
(blue online) represents the initial position, the light-grey line
(green online) the position of the droplet at time $t_1^* = 0.1$.
The arrows depict the velocity of the flow in the fixed reference
frame.}
 \label{Fig:Prescribed sigma = sin}
\end{figure}

Several swimming microorganisms as well as artificial microswimmers can be described
through the "squirmer" model \cite{lig52}, starting from a prescribed flow field at the interface.
In the present simulation, we reproduce this standard behavior without directly prescribing a surface flow,
but a surface tension distribution instead. In particular, we considered the case a spherical
droplet of radius $R_0$ and we derived the velocity field originating from a
non-uniform surface tension distribution. To this aim, we disregarded the surfactant fields both in the bulk
and in the surface, and we prescribed a surface
tension field $\sigma(\theta) = \sigma_0 (1 + \epsilon
\sin\theta/2)$, where $\theta$ was the angular position with
respect to the radial axes, in the counterclockwise direction. The
characteristic scales are $L_0 = R_0$ for the
length and $t_0 = \sqrt{\pi R_0^3 \rho / 2\epsilon \sigma_0}$ for
the time, as $t_0$ gives an estimate of the time required to
produce a displacement $R_0$ due to Marangoni flow. In
\Fig{Prescribed sigma = sin}, the black line (blue online)
represents the initial position of the droplet, the light-grey
line (green online) represents the contour of the droplet, almost
undeformed, at time $t_1^* = 0.1$. The arrows depict the velocity
field $\textbf{u}$ in the fixed frame of reference, i.e. in the lab
frame, at time $t^*_1$. The exhibited behavior is typical of the
so-called "neutral squirmers" \cite{lig52}, as the fluid is pushed away from the
droplet at the leading edge (in the front), and is
pulled towards the droplet at the trailing edge (in the back).

\begin{figure}
\centering
\includegraphics[width=0.65\textwidth]{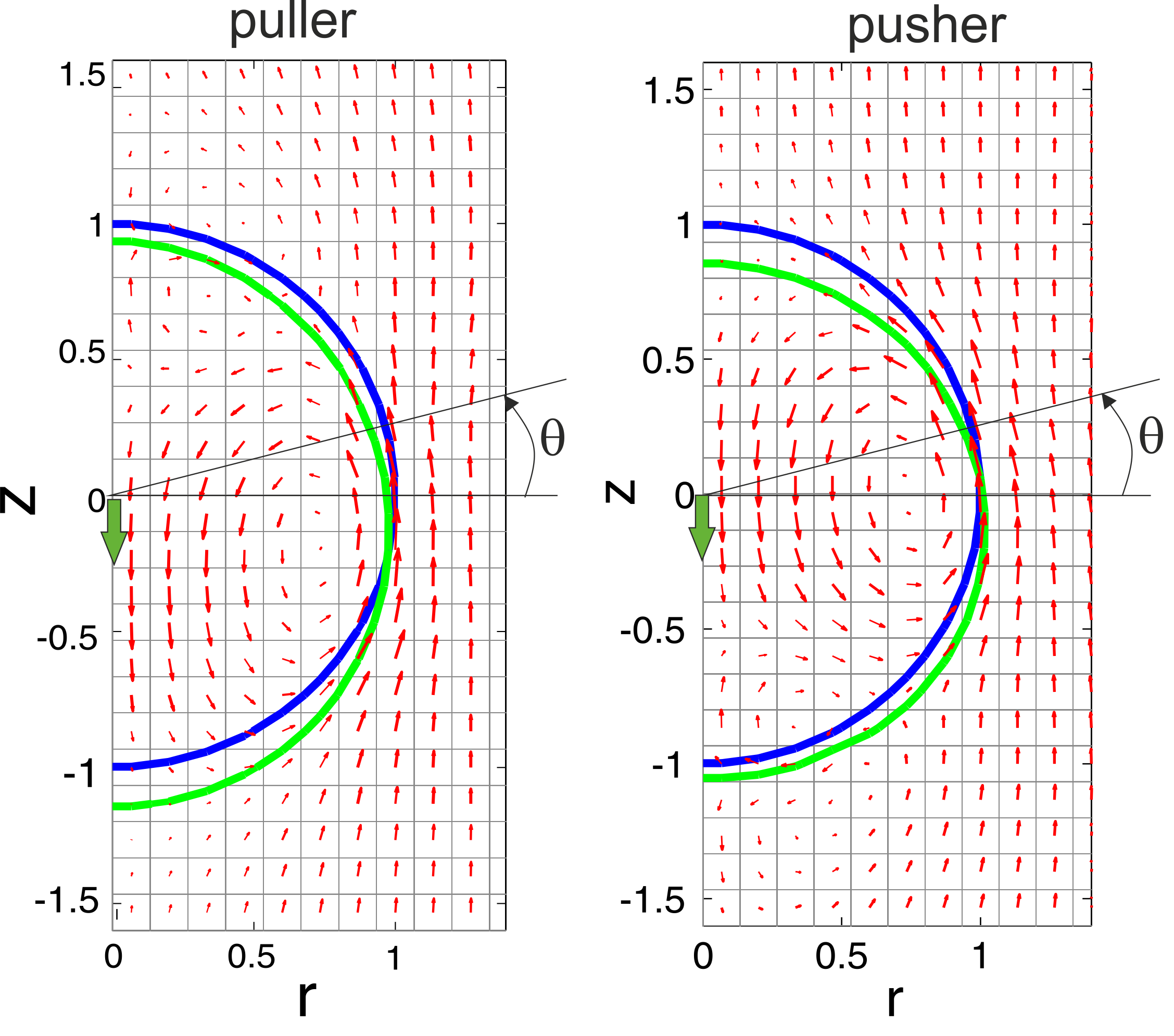}
\caption{Numerical simulation of a droplet moving by Marangoni
flow in the downward direction, when a surface tension field is
prescribed as $\sigma(\theta) = \sigma_0 (1 + \epsilon
\textmd{cos}\theta \pm \sin (2\theta))$. The sign $+$ corresponds
to the figure on the left, the $-$ to the figure on the right.
Parameters: $R_0 = 25 \mum$, $\epsilon = 0.4$. The blue lines
represent the initial position, the green lines the position of
the droplet at time $t^* = t / t_0 = 0.12$. The thin arrows (red online) depict
the velocity of the flow in the co-moving frame, shifting with the
droplet. The two thick arrows (green online) indicate the direction of the droplet's motion.
The two droplets reproduce the typical puller (left),
pusher (right) behavior exhibited by squirmers.}
 \label{Fig:Pusher/puller}
\end{figure}

In \Fig{Pusher/puller} we show that, by prescribing a surface
tension field $\sigma(\theta) = \sigma_0 [1 + \epsilon \cos\theta
\pm \sin (2\theta)]$, it is possible to reproduce the typical
behavior of squirmers \cite{lig52} acting as "pullers" (left figure,
corresponding to the sign $+$) and "pushers" (right figure,
corresponding to the sign $-$). At the droplet's trailing and leading edge
the fluid velocity has indeed inward orientation respect to the
interface, for the puller, outward  for the pusher. The black
lines (blue online) display again the initial position, the
light-grey lines (green online) the position of the droplet at
time $t_1^* = 0.1$. The arrows represent the velocity field in the
co-moving frame of reference, displacing with the droplet. These
velocities were calculated by subtracting the average velocity of
the droplet, as derived by \Eq{Droplet's velocity, numerical},
from the velocity field $\textbf{u}$.

\clearpage

\subsection{Insoluble surfactants, advection and diffusion}
\begin{figure*}
\centering
\includegraphics[width=0.8\textwidth]{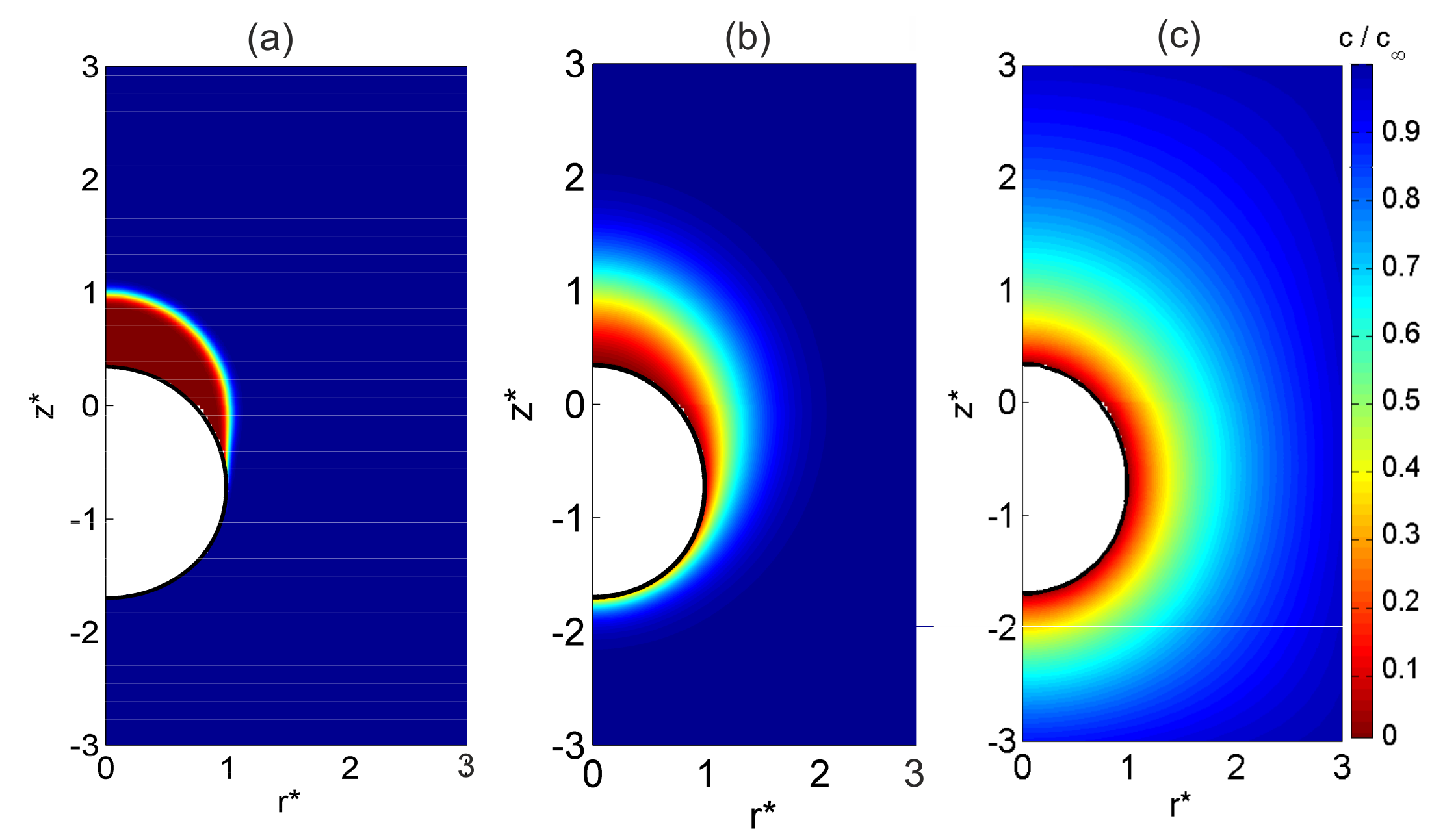}
\caption{Evolution of an initial bulk concentration field $c(t =
0,\mathbf{x})  = c_\infty H(\Phi)$ around a moving droplet, here depicted
at time $t_1^* = 0.8$ 
for different values of the bulk diffusivity
constant $D$. The motion of the droplet is prescribed and equal to
the one described in \Fig{Prescribed sigma = sin}. No coupling
between the surface and the bulk is present (insoluble
surfactants). (a) Low diffusive constant, advection dominates ($D = 10^{-8} m^2/s$)
(b) Diffusion and advection are both relevant ($D = 8.5 \cdot 10^{-7} m^2/s$)
(c) High diffusive constant, diffusion dominates ($D = 10^{-4} m^2/s$). }
 \label{Fig:Bulk conc moving drop}
\end{figure*}

In the following simulation, we consider the case of a droplet moving by Marangoni flow,
with two non-interacting concentration fields, one covering the surface, $\Gamma$, the other, $c$, dissolved in the surrounding liquid.
From the physical point of view, such a situation can occur when an insoluble surfactant
covers the surface and a second specie is dissolved in the bulk, in equilibrium
condition with the content of the droplet (e.g. liquid crystals in the emulsion droplets of \cite{thu12}).
This test case tackled the advection of the concentration fields, due to the motion of the droplet itself.
Since the surfactants were taken as insoluble, no exchange with the bulk was
present and $j = 0$ in \Eq{surface conc adv-diff eqn}.
We therefore examined the evolution of the concentration fields, for a spherical droplet
of initial radius $R_0$ moving by Marangoni flow. The motion was externally
prescribed, by imposing a surface tension field
$\sigma(\theta) = \sigma_0 (1 + \epsilon \sin\theta/2)$, which gave origin to
a velocity profile like in \Fig{Prescribed
sigma = sin}. Thus, for this test case, the retroaction of
the surfactant concentration on the surface tension \Eq{Langmuir eqn}
was disregarded. The initial bulk concentration field was taken as
$c = c_{\infty} H(\Phi)$, while the initial surface concentration was uniform
and equal to $\Gamma_0$.
The characteristic scales of the problem were $L_0 = R_0$ for the
length, $t_0 = \sqrt{\pi R_0^3 \rho / 2\epsilon \sigma_0}$ for
the time and $\Gamma_0$ for the surface concentration. The concentration fields
moved due to both advection and diffusion. By changing the diffusion constants $D$ and $D_s$,
which give an indication of the speed of the diffusive process,
and keeping all the other parameters the same, different behaviors
of the concentration fields were expected. To asses the effect of
$D$ on the time evolution of the bulk concentration field $c$, we
prescribed an initial stepwise uniform bulk concentration equal to
$c_\infty$ for $r>R_0$, $c_R$ for $r=R_0$. We then examined the
bulk concentration field after some time, for
different values of $D$. In \Fig{Bulk conc moving drop}.a advection
dominates over diffusion ($D = 10^{-8} m^2/s$). In this case the
dissolved species is mostly advected by the velocity field (see Fig.~\ref{Fig:Prescribed sigma = sin}),
but it does not have the time to diffuse. Therefore a neat wake is
formed behind the droplet. In \Fig{Bulk conc moving drop}.b, advection
and diffusion are both relevant ($D = 8.5 \cdot 10^{-7} m^2/s$):
the final distribution in asymmetric respect to the equatorial
plane of the droplet but a consistent smoothening of the initial field appears.
In \Fig{Bulk conc moving drop}.c bulk diffusion dominates over advection
($D = 10^{-4}m^2/s$): the initial distribution of the dissolved species is smoothed
out but still symmetric respect to the equatorial plane of the droplet,
regardless of the fluid velocity field.

To assess the effect of the surface diffusivity $D_s$ on the time
evolution of the surface concentration field $\Gamma$, we adopted the
same procedure, by prescribing an initial sinusoidal surface concentration field
$\Gamma = \Gamma_0 (1 + \epsilon \sin \theta) $ with $\epsilon = 1/2$ and examining
the effects as the time passed, for different values of
$D_s$. In \Fig{Surface adv-diffusion, different Ds}.a surface diffusion appears to be
dominant over advection ($D_s = 10^{-3} m^2/s$): the initial
surface concentration (black line, blue online) is uniformly
leveled and evolves in a purely diffusive fashion (light-grey line, green online), regardless of the
fluid's motion. In \Fig{Surface adv-diffusion, different Ds}.b-c
($D_s = 10^{-5} m^2/s$ and $D_s = 10^{-7} m^2/s$ respectively), advection
is also relevant and the evolution of the surface concentration
differs from the purely diffusive one. In these
cases, surfactants are piled up at $\theta = \pi/4$ and $\theta = -\pi/2$,
following the motion of the fluid along the interface (see the
direction of the vortices in \Fig{Prescribed sigma = sin}, for a
droplet moving in the downwards direction).
\begin{figure}
\centering
\includegraphics[width=0.8\textwidth]{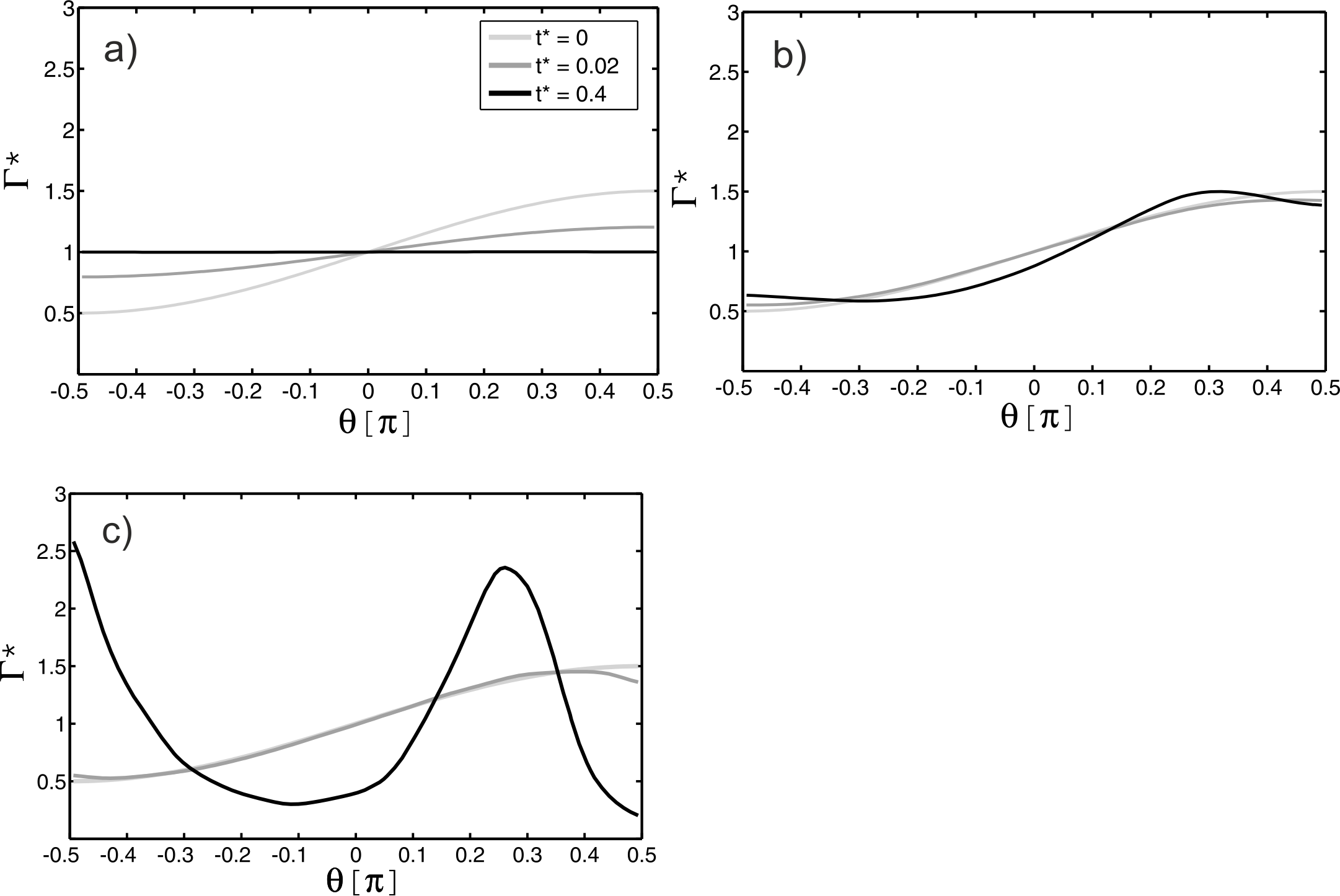}
\caption{Time evolution of an initial distribution of insoluble
surfactants $\Gamma$ (light-grey line) for different values of the
surface diffusivity: (a) $D_s
= 10^{-3} m^2/s$ (b) $D_s = 10^{-5} m^2/s$ (c) $D_s= 10^{-7} m^2/s$.
The droplet is moving, due to a prescribed
surface tension distribution, which induces a motion like the
one depicted in \Fig{Prescribed sigma = sin}. The different curves
correspond to different time instants, as indicated in the legend.
}
 \label{Fig:Surface adv-diffusion, different Ds}
\end{figure}

\clearpage

\subsection{BZ-reactions driven droplets}
\label{sec:Results, BZ reactions} In these simulations, we
considered a droplet with BZ reactions taking place on its surface, of the kind described in \cite{thu12}.
In particular, we wanted to assess the inception of a self-sustained
propulsion mechanism, as a consequence of an initial perturbation (impulse), thus validating the
simplified numerical treatment developed in Sec.~\ref{sec:Numerical model, chemical
reactions}. In order to mimic this scenario, we adopted the following procedure.
The motion of the droplet was artificially initiated by
prescribing a surface tension distribution $\sigma(\theta) =
\sigma_0 (1 + \epsilon \sin\theta/2)$, giving origin to a
downwards oriented motion (see \Fig{Prescribed sigma = sin}).
The concentration fields were allowed to evolve according to their
coupled advection-diffusion equations. However, for some time,
the surface tension was still prescribed, thus acting as an
external forcing, rather than being determined from the surface
concentration of fresh surfactants.
After some time, the artificial driving was switched off and, from that moment on, the
surface tension distribution was derived by Langmuir equation
(\ref{Eq:Langmuir eqn}). We observed that, after the removal of
the forcing, the droplets kept moving, eventually reaching a
non-null steady-state velocity. Hence, the initial perturbation
initiated a self-sustained motion. In \Fig{Velocity center,
driving} we show the time evolution of the average velocity of the
droplet, calculated by \Eq{Terminal velocity, Marangoni test}. The velocity
is negative because the droplet is moving in the
downward direction, along the z-axes. If the external forcing is
not interrupted, the droplet keeps accelerating smoothly (grey
line). If the external forcing is interrupted (black line) at time
$t^*_s$, the droplet decelerates, but does not stop: the
surfactants have self-organized in such a way that the motion
continues, eventually reaching a steady-state velocity $U_0^*$.

\begin{figure}
\centering
\includegraphics[width=0.6\textwidth]{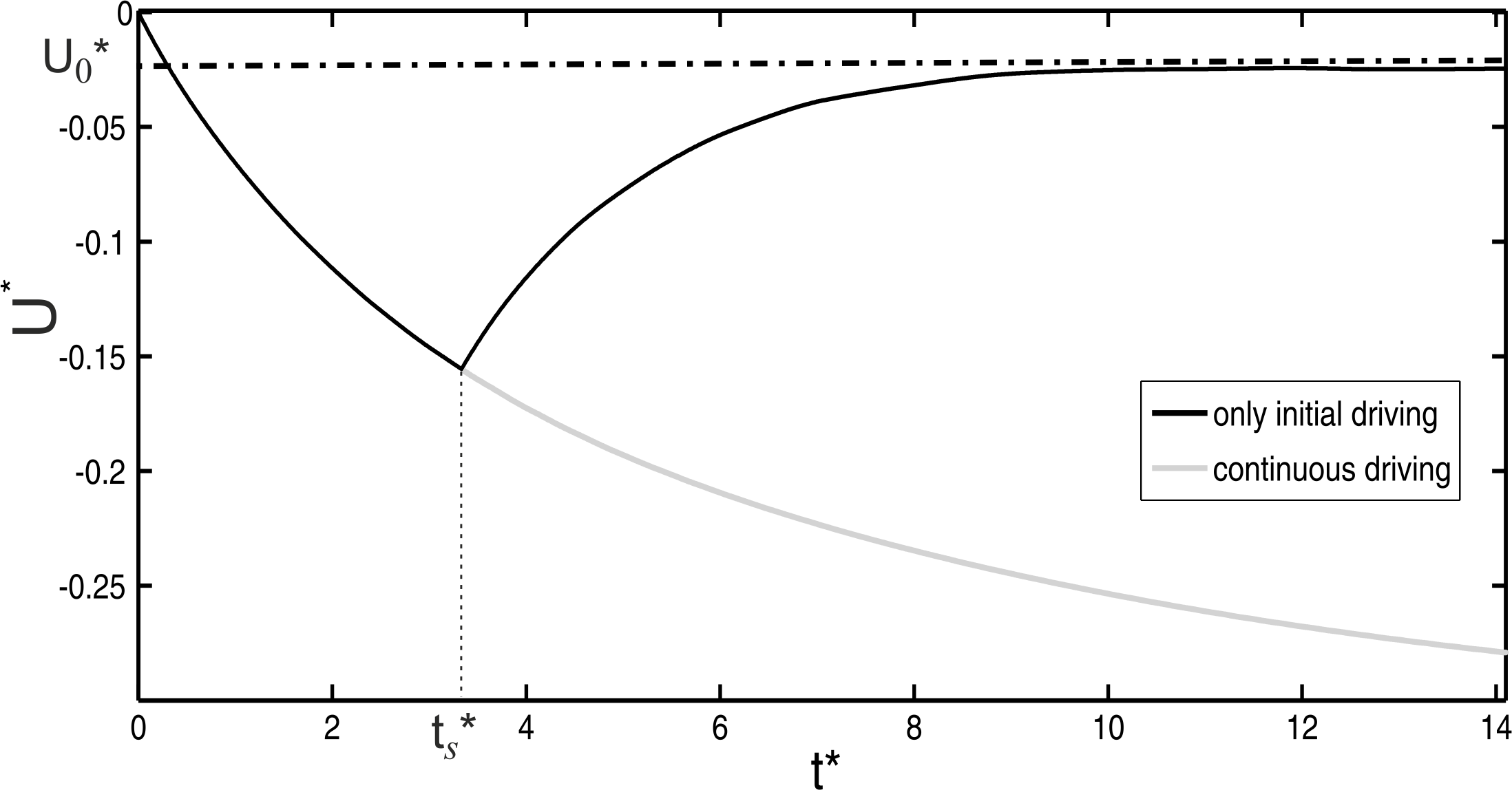}
\caption{Time evolution of the velocity of a droplet propelled by
BZ-reactions. The motion is artificially initiated by prescribing
a surface tension distribution like the one used in
\Fig{Prescribed sigma = sin}. The grey line represents a case
where the external driving is preserved and the droplet keeps
accelerating. The black line represents a case where the driving
is turned off at time $t_s^* = 3.4$. The droplets decelerates, but
it eventually reaches a steady-state velocity $U_0^* = -0.2$.
Parameters: $D = 10^{-10} m^2/s$, $D_s = 10^{-9} m^2/s$, $R_0 =
6\mum$, $r_a = r_d = 1000$, $r_c = 10$.}
 \label{Fig:Velocity center, driving}
\end{figure}

In \Fig{Bulk conc coupling, moving droplet} we show the time
evolution of the bulk concentration field, with realistic parameters for the
diffusive constants $D = 10^{-10} m^2/s$ and $D_s = 10^{-9}
m^2/s$. The droplet is moving in the downward direction. We notice
that, in the wake of the droplet, there is an
area where a lower surfactant concentration is present.

In \Fig{Surf conc 2 species, moving droplet}.a the surface concentration fields
of fresh surfactant $\Gamma^*_f$ (solid lines), and waste surfactant
$\Gamma^*_w$ (dashed lines), are depicted, at different instants.
\Fig{Surf conc 2 species, moving droplet}.b displays the corresponding surface tension distributions.
A gradient of fresh surfactant, thus of surface tension, originates as a consequence of the initial driving.
As time passes, such a gradient persists, originating the self-propulsion mechanism of the droplet.
\begin{figure}
\centering
\includegraphics[width=0.95\textwidth]{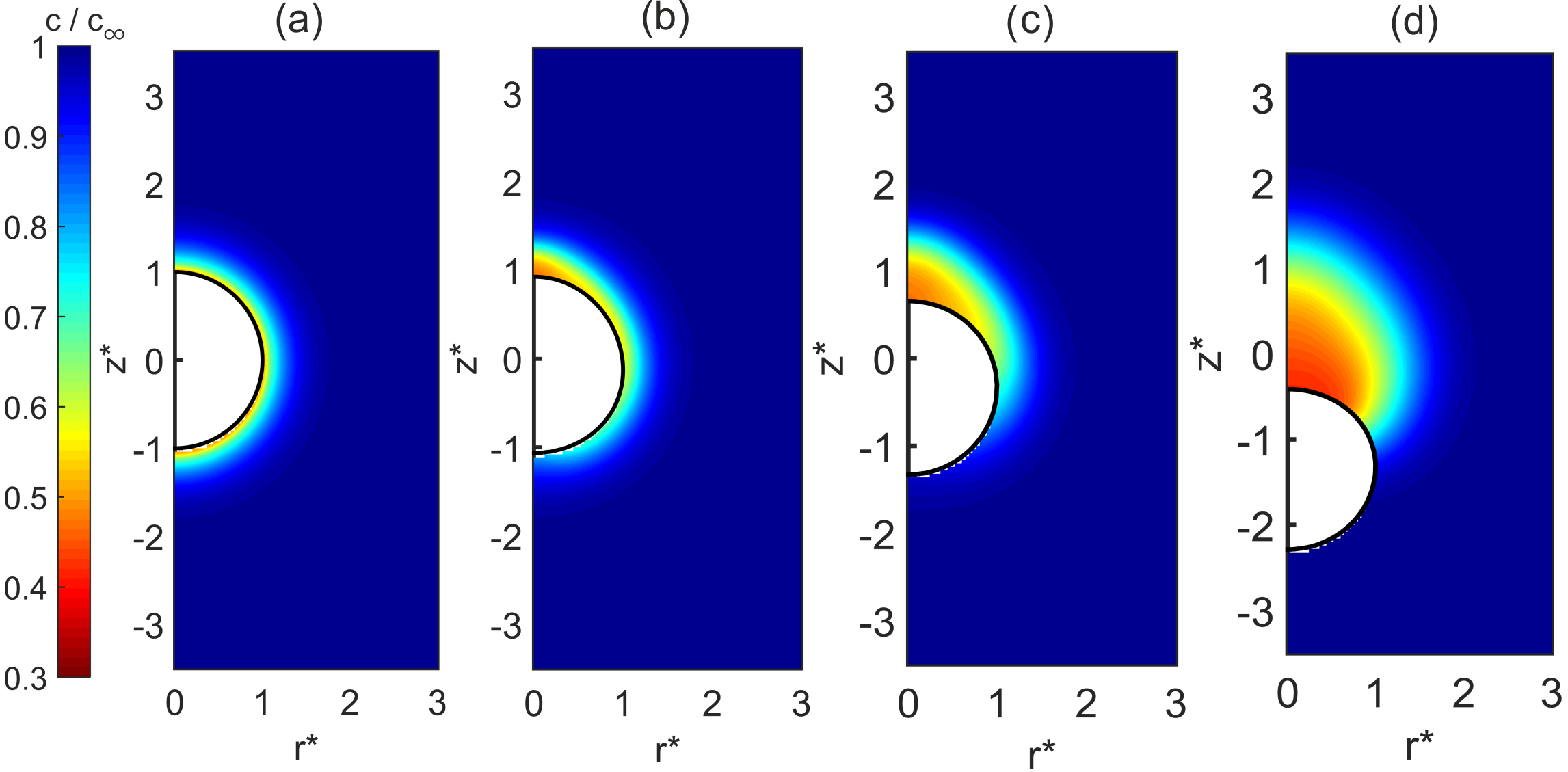}
\caption{Bulk concentration field of fresh surfactants $c$ around
the BZ-reactions driven droplet of \Fig{Velocity center, driving},
moving downwards, at different times (a) $t^* = 0$ (b) $t^* = t_s^* = 3.4$, when the external forcing
is turned off (c) $t^* = 6.2$ (d) $t^* = 12.5$. Parameters: as in \Fig{Velocity center, driving}.
A decumulation area of the surfactant appears in the wake of the droplet.}
 \label{Fig:Bulk conc coupling, moving droplet}
\end{figure}

\begin{figure}
\centering
\includegraphics[width=0.95\textwidth]{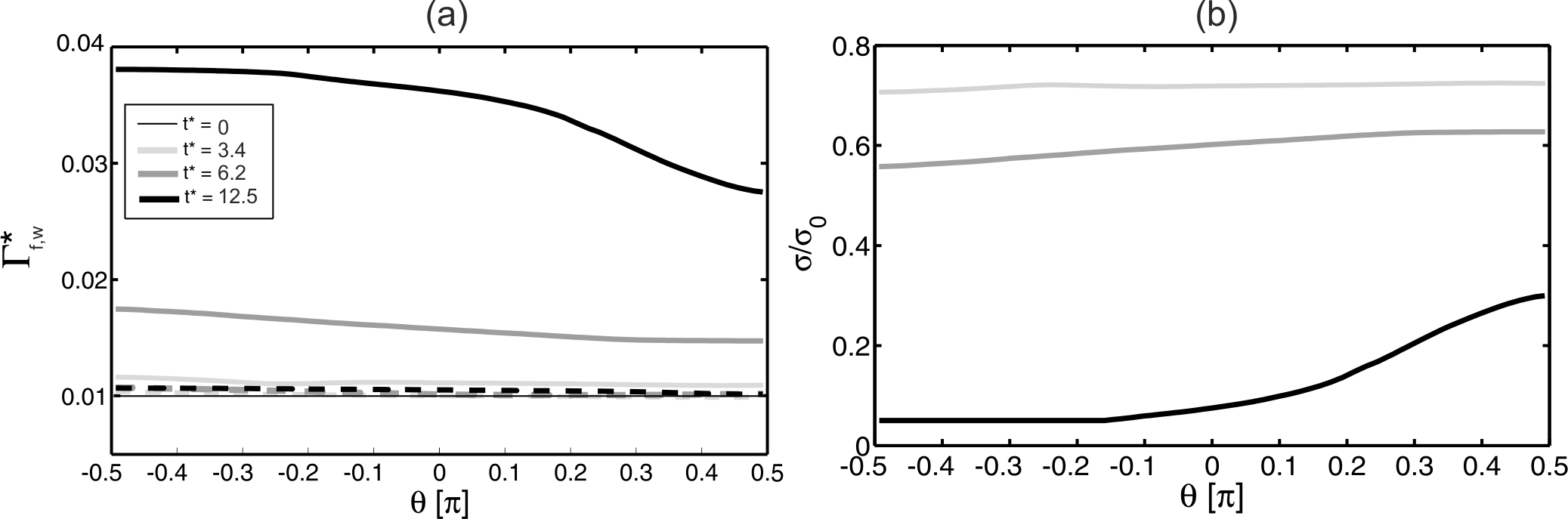}
\caption{(a)Surface concentration field of fresh surfactants $\Gamma^*_f$ (solid lines) and
waste surfactants $\Gamma^*_w$ (dashed lines) and (b) surface tension for a BZ-reactions
driven droplet. The different colours represent different times $t^* = 0$ (thin black lines) $t^* = 3.4$
(light-grey lines) $t^* = 6.2$ (dark-grey lines) $t^* = 12.5$ (black lines).
Parameters: as in \Fig{Velocity center, driving}.}
 \label{Fig:Surf conc 2 species, moving droplet}
\end{figure}

\subsection{Chemotaxis}
\label{sec:Results, chemotaxis}
The purpose of the present simulations was to reproduce a typical chemotactic behavior,
where a microswimmer moves along a gradient of chemical concentration and
eventually slows down due to the inception of surface saturation.
To this aim, we considered a spherical droplet of radius $R_0$ with
adsorption/desorption of surfactant at the interface (soluble surfactants), but no
chemical reactions. The droplet was introduced in a channel of
length $L$, where a linear gradient of bulk concentration $c$ was
imposed \beq c(z) = c_{\infty} \frac{z}{L} \label{Eq:Bulk conc
field, chemotaxis}\eeq with $c_{\infty}$ a constant. The droplet
was expected to start moving in the upward direction, following
the gradient, mimicking the behavior of several biological and
artificial microswimmers. In \Fig{Chemotaxis, average velocity} we
display the time evolution of the average velocity of the droplet,
as calculated by \Eq{Terminal velocity, Marangoni test}. The
droplet immediately started moving in the expected direction
(upwards, along the gradient), accelerating. The surfactant was
adsorbed at the surface, thus reducing the surface tension until
this reached, first only in some parts (at time $t_{\sigma}^*$),
then everywhere, its minimum possible value $\epsilon_{\sigma}$,
according to \Eq{Langmuir eqn, cutoff}. Therefore after time
$t_{\sigma}^*$, the surface concentration started to decelerate
due to the inception of saturation.

\begin{figure}
\centering
\includegraphics[width=0.7\textwidth]{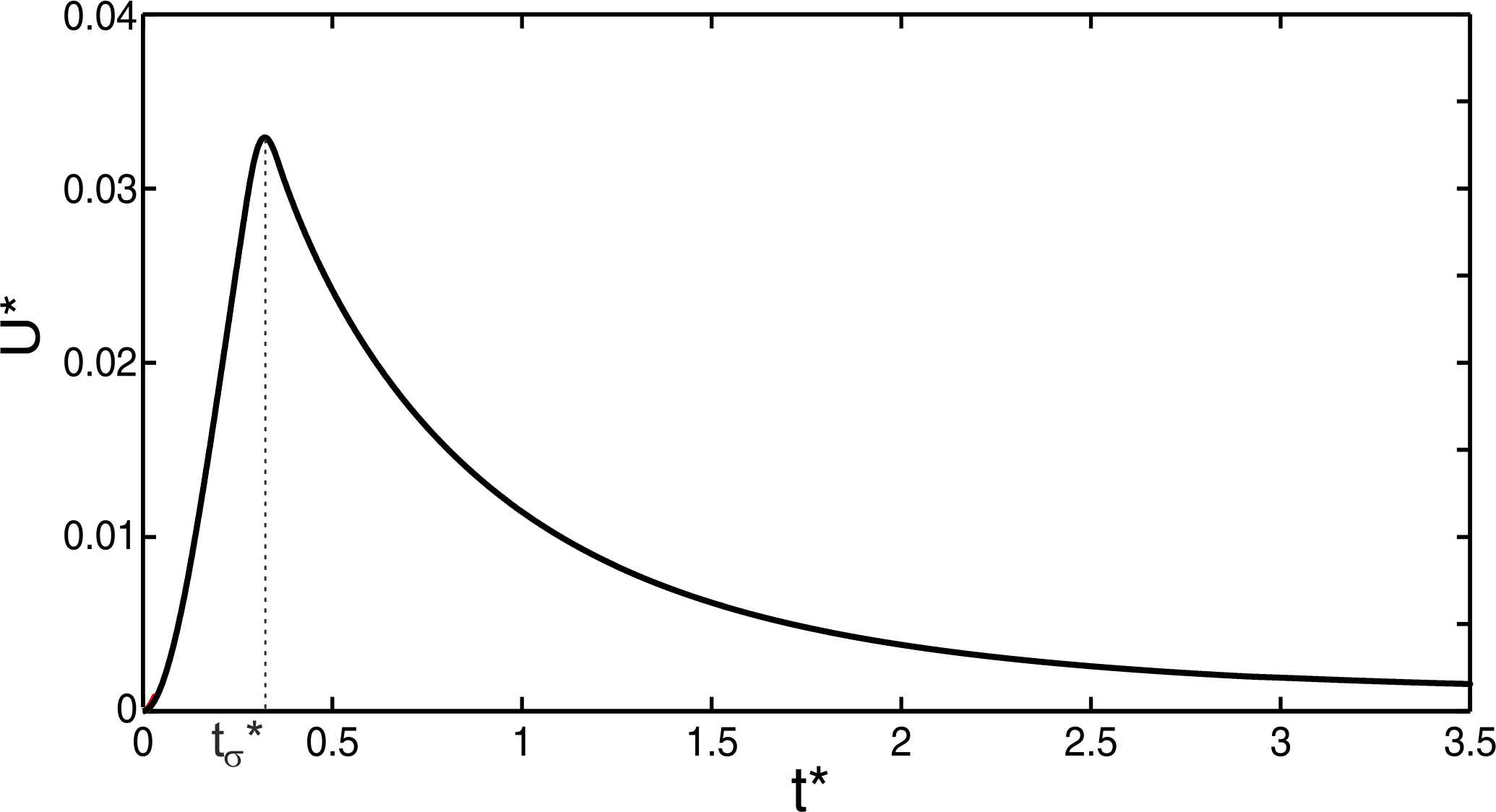}
\caption{Time evolution of the average velocity of a droplet with
adsorption-desorption of surfactant, immersed in a linear gradient
of surfactants in the bulk \ref{Eq:Bulk conc
field, chemotaxis}. The droplet starts to accelerate along the gradient, in
the upward direction, but eventually slows
down, starting from time $t_{\sigma}^*$, due to the inception of
surface saturation. Parameters:$R_0 = 6\mum$, $L = 60 \mum$,
$\beta_s = 1$, $r_a = 100$, $r_d = 10$, $c_\infty = 10^{-6}
mol/m^3 $.}
 \label{Fig:Chemotaxis, average velocity}
\end{figure}

\section{Conclusions} In this paper we considered the
numerical study of artificial microswimmers, consisting of
microdroplets moving by Marangoni flow, such as liquid crystal
emulsion droplets \cite{ped12} and BZ-reactions-driven droplets
\cite{thu12}. A detailed description of such systems
lays beyond the scope of the present work, their complexity arising from
the coexistence of an hydrodynamic motion, adsorbtion/desorption
of surfactants at the interface, as well as reactions of different species, so
that one has to deal with mutually interacting concentration
fields. The aim of the present work was to provide a numerical
treatment for the main phenomena playing a role in the above-mentioned systems.
We developed a second-order level-set model based on a continuous
surface force approach. The main elements of the model are mass
and momentum conservation, advection-diffusion of concentration
fields both in the bulk (liquid surrounding the droplet) and on
the droplet's surface. For the interaction between the bulk and
the surface we considered several cases: a. the active exchange of
surfactants between the bulk and the surface b. a concentration
field dissolved in the liquid and in equilibrium with the content
of the droplet (applicable to the behavior of liquid crystals
based systems \cite{thu12}). We also implemented a simple
treatment of BZ reactions depleting surfactant at the surface. The
numerical results were validated by comparison with analytical
test cases, showing a second-order convergence respect to the grid
size. With the present model, we reproduced the typical
pusher/puller behavior of "squirmers". We presented a possible
mechanism for self-propulsion in BZ-reaction driven droplets, by
artificially initiating the motion by means of an impulse. After the
external forcing stopped, the droplet continued to move reaching a non-zero
steady-state speed. We also reproduced a typical chemotactic behavior, where a
microswimmer moves along a gradient of chemical concentration and
eventually slows down due to the inception of surface saturation.

\section{Acknowledgments} The research leading to these results has
received funding from the People Programme (Marie Curie Actions)
of the European Union's Seventh Framework Programme FP7/2007-2013/
under REA grant agreement n$^{\circ}$ [628154]. The author would
like to express her gratitude to J\"{u}rgen Vollmer, James
Sethian, Robert Saye, Andrea Prosperetti and Stephan Herminghaus
for insightful discussions and feedback on the manuscript. She
would also like to acknowledge the hospitality at the Lawrence
Berkeley National Laboratories (Berkeley, CA), where part of this
research has been carried out.


\end{document}